\documentclass[fleqn,usenatbib]{mnras}

\usepackage{newtxtext,newtxmath}

\usepackage[T1]{fontenc}

\DeclareRobustCommand{\VAN}[3]{#2}
\let\VANthebibliography\thebibliography
\def\thebibliography{\DeclareRobustCommand{\VAN}[3]{##3}\VANthebibliography}

\usepackage{graphicx}	
\usepackage{amsmath}	
\usepackage{times}
\usepackage{xcolor}  

\newcommand{\HII}{H\,{\sc ii}\rm\,}
\newcommand{\NII}{{[N\,{\sc ii}]}}
\newcommand{\NIIs}{{[N\,{\sc ii}]\,}}
\newcommand{\SII}{{[S\,{\sc ii}]}}
\newcommand{\SIIs}{{[S\,{\sc ii}]\,}}
\newcommand{\SIII}{{[S\,{\sc iii}]}}
\newcommand{\SIIIs}{{[S\,{\sc iii}]\,}}
\newcommand{\OIII}{{[O\,{\sc iii}]}}
\newcommand{\OIIIs}{{[O\,{\sc iii}]\,}}
\newcommand{\OII}{{[O\,{\sc ii}]}}
\newcommand{\OIIs}{{[O\,{\sc ii}]\,}}
\newcommand{\OI}{{[O\,{\sc i}]}}
\newcommand{\OIs}{{[O\,{\sc i}]\,}}
\newcommand{\FeII}{{[Fe\,{\sc ii}]\,}}
\newcommand{\Ha}{H$\alpha$}
\newcommand{\Has}{H$\alpha$\,}

\newcommand{\ZLS}{$Z_\text{Te; LS12}$ }
\newcommand{\ZYates}{$Z_\text{Te; Y20}$ }
\newcommand{\ZSII}{$Z_\text{Te; SII}$ }
\newcommand{\ZLSe}{$Z_\text{Te; LS12}$}
\newcommand{\ZYatese}{$Z_\text{Te; Y20}$}
\newcommand{\ZSIIe}{$Z_\text{Te; SII}$}

\defcitealias{Yates20}{Y20}

\title[Resolved direct metallicity in SAMI]{Spatially resolved direct method metallicity in a high-redshift analogue local galaxy: temperature structure impact on metallicity gradients}

\author[A. J. Cameron et al.]{
Alex J. Cameron,$^{1,2}$\thanks{E-mail: alexc@student.unimelb.edu.au}
Tiantian Yuan,$^{3,2}$
Michele Trenti,$^{1,2}$
David C. Nicholls,$^{4,2}$\newauthor
and Lisa J. Kewley$^{4,2}$
\\
$^{1}$School of Physics, The University of Melbourne, Parkville, VIC 3010, Australia\\
$^{2}$ARC Centre of Excellence for All-Sky Astrophysics in 3 Dimensions (ASTRO 3D), Australia\\
$^{3}$Centre for Astrophysics and Supercomputing, Swinburne University of Technology, Hawthorn, Victoria 3122, Australia\\
$^{4}$Research School of Astronomy and Astrophysics, The Australian National University, Cotter Road, Weston, ACT 2611, Australia\\
}

\date{Accepted XXX. Received YYY; in original form ZZZ}

\pubyear{2020}

\begin{document}
\label{firstpage}
\pagerange{\pageref{firstpage}--\pageref{lastpage}}
\maketitle

\begin{abstract}
We investigate how H{\sc ii} region temperature structure assumptions affect ``direct-method'' spatially-resolved metallicity  observations using multispecies auroral  lines in a galaxy from the SAMI Galaxy Survey. SAMI609396B, at redshift $z=0.018$, is a low-mass galaxy in a minor merger with intense star formation, analogous to conditions at high redshifts. We use three methods to derive direct metallicities and compare with strong-line diagnostics.
The spatial metallicity trends show significant differences among the three direct methods.
Our first method is based on the commonly used electron temperature $T_e$([O{\sc iii}]) from the [O{\sc iii}]$\lambda$4363 auroral line and a traditional $T_e$([O{\sc ii}]) -- $T_e$([O{\sc iii}]) calibration. The second method applies a recent empirical correction to the O$^+$ abundance from the  [O{\sc iii}]/[O{\sc ii}] strong-line ratio.  The third method infers the $T_e$([O{\sc ii}]) from the [S{\sc ii}]$\lambda\lambda$4069,76  auroral lines.
The first method favours a positive metallicity gradient along SAMI609396B, whereas the second and third methods yield flattened gradients. Strong-line diagnostics produce mostly flat gradients, albeit with unquantified contamination from shocked regions. We conclude that overlooked assumptions about the internal temperature structure of H{\sc ii} regions in the direct method can lead to large discrepancies in metallicity gradient studies.
Our detailed analysis of SAMI609396B underlines that high-accuracy metallicity gradient measurements require a wide array of emission lines and improved spatial resolutions in order to properly constrain excitation sources, physical conditions, and temperature structures of the emitting gas.
Integral-field spectroscopic studies with future facilities such as \emph{JWST}/NIRSpec and ground-based ELTs will be crucial in minimising systematic effects on measured gradients in distant galaxies.

\end{abstract}




\section{Introduction}

Abundances of heavy elements (metallicities) in the interstellar medium (ISM) of galaxies are enriched by stellar nucleosynthesis and trace star formation histories and gas-flow processes that ultimately shape the galaxy population.
In particular, the spatial distribution of metallicity offers a powerful probe on the role of mergers, outflows, gas mixing, and gas accretion in transforming galaxies \citep[e.g.,][]{Edmunds95,Kewley10,Torrey12,Magrini16,Finlator17,Ma17, Bresolin19,Tissera19,Hemler20}.
Spatial distributions of metals are often summarised as radial abundance gradients and azimuthal variations \citep[e.g.,][]{Searle71,Vila-Costas92,Li13,Ho15,Ho19}, with both
negative and flat metallicity gradients widely observed in the Milky Way and other local galaxies \citep[e.g.,][]{Deharveng00, Bresolin04, Berg13, Berg20}.

Spatially resolved studies of galaxies are now far more accessible compared to a decade ago, thanks to the advent of integral-field unit (IFU) spectroscopy.
Multiplexed IFU surveys (e.g. CALIFA \citealt{Sanchez12}; SAMI \citealt{Bryant15}; MaNGA \citealt{Bundy15}) have afforded large samples of gradient measurements in the local Universe. Studies find a dependence on stellar mass:  low-mass galaxies ($\sim$10$^9$ $M_\odot$) show almost flat gradients, with negative gradients steepening to high masses \citep{Belfiore17, Poetrodjojo18}.
Spatially resolved measurements become more challenging at high redshift and observations show a substantial amount of scatter \citep{Yuan11, Jones13, Leethochawalit16, Carton18, Wang19b, Curti20b}. One major caveat in using this broad range of observations  to develop a coherent model of galaxy evolution is that different measurement techniques are often in disagreement.

A number of observational methods exist for determining the oxygen abundance (metallicity hereafter) of the ISM  in galaxies from emission line spectroscopy (see \citealt{MaiolinoMannucci19, Kewley19} for recent reviews).
However, different techniques often show large offsets up to 0.7 dex \citep[e.g.][]{KewleyEllison08,Peimbert17}.
This stark disagreement between different metallicity measurement techniques presents an ongoing challenge for studying chemical evolution of galaxies.

Emission line strengths in the photoionised nebulae around hot O- and B-type stars (\HII regions) are sensitive to electron temperature ($T_e$), in addition to ionic abundances, ionisation parameter, and ISM pressure. Thus, a desirable approach to metallicity measurement is to use ratios of auroral emission lines and corresponding strong nebular emission lines to explicitly determine $T_e$, and subsequently metallicity (Direct Method; e.g., see \citealt{PerezMontero17} for an overview).
This ``direct method'' is traditionally considered the gold standard in abundance determination \citep[e.g.][]{MaiolinoMannucci19}, and underpins the calibration of many alternative techniques \citep[e.g.][]{PettiniPagel04, Curti20a}.
However, one major practical issue with the direct method is that the faintness of the optical auroral lines severely limits its application. An alternative $T_e$-based method outlined by \citet{Jones20} determines oxygen abundance based instead on far-infrared oxygen lines (\OIIIs52\micron\, or \OIIIs88\micron). This is expected to be favourable beyond $z\gtrsim5$ where these far-IR features can be observed with millimeter instruments such as \emph{ALMA}, but is difficult to apply at lower redshifts.

Due to the faintness of auroral lines required for the direct method, strong-line methods are widely adopted in observations. Strong-line methods use ratios of the brightest rest-frame ultra-violet and optical emission lines to empirically determine the metallicity with calibrations based on either  direct-method observations \citep[e.g.][]{PettiniPagel04, Pilyugin05, Curti20a} or stellar population synthesis and photoionisation models \citep[e.g.][]{Kewley02, KobulnickyKewley04, Dopita16}. Strong-line methods vastly expand the redshift and mass range of galaxies for which metallicities can be derived.
However, it has been widely observed that metallicities measured with different methods often disagree \citep[e.g.][]{KewleyEllison08, Moustakas10, MoralesLuis14}.
In particular, theoretical methods, are reliant on simple geometries, such as spherical or plane parallel, and assume a constant temperature, constant density, or a constant pressure.

Despite the baseline role of the direct method, it does have limitations beyond practical detection-rate issues \citep{Nicholls20, Yates20}.
\HII regions are complex structures and summarising their conditions with integrated measurements of emission line ratios carries many assumptions.
For example, \HII regions are known to have internal temperature variations \citep{Peimbert67, Kewley19}. An observed emission line ratio samples the luminosity-weighted average conditions of the emitting nebulae \citep{Nicholls20}.
The direct method is best applied by constructing a multi-zone temperature model using auroral lines from multiple ionic species \citep[e.g.][]{PerezMontero17, Berg20}. Commonly used auroral lines include those from O$^{2+}$, O$^{+}$, N$^{+}$ or S$^{2+}$ ions.\footnote{The Cl$^{2+}$ and Ar$^{3+}$ ions can provide similar temperature probes to complement O$^{2+}$ measurements, however are usually too faint to be detectable.} This allows internal temperature gradients to be sampled since ions with differing ionisation energies preferentially sample different sub-regions of the nebulae.

However, measuring auroral lines from multiple species in observations presents a difficult practical challenge. Even detection of a single auroral line, commonly \OIIIs$\lambda$4363, is generally considered a favourable outcome.
But since the \OIIIs$\lambda$4363 line is only produced in the hottest regions of a nebula, a resulting $T_e$-derived metallicity may be a lower limit to the true metallicity if there is a temperature gradient \citep{Kewley19}.
To overcome the lack of direct constraints on the multi-zone temperature structure, abundance measurements are often made adopting empirical relations between temperatures from different ions. For example, the \OIIs temperature ($T_e$(\OII)) is indirectly inferred from the  \OIIIs temperature ($T_e$(\OIII); based on \OIIIs$\lambda$4363) using the  $T_e$(\OII) -- $T_e$(\OIII) relation (e.g. \citealt{Izotov06, LopezSanchez12, PerezMontero17}).
Recently, \citet{Yates20} show that at low O$^{2+}$/O$^{+}$, this approach can lead to large deficits in the measured O$^{+}$ abundance, causing total oxygen abundances to be underestimated by up to $\sim$0.6 dex.

Studying metallicity in spatially resolved detail exacerbates the practical limitations of the direct method. Indeed, direct method metallicities have been mapped only for the Milky Way and small samples of large nearby spiral galaxies \citep{Deharveng00, Bresolin04, Li13, Berg13, Berg15, Berg20, Croxall15, Croxall16, Ho19}, exploring only a very narrow subset of the galaxy population.
Here we leverage public release IFU data from the SAMI Galaxy Survey to expand spatially resolved direct method metallicity measurements to a new parameter space. From a search of auroral lines in SAMI Data Release 2 data cubes,  we identify one particularly strong candidate: SAMI609396. This target is a minor-merger system and one galaxy in the system (SAMI609396B) is experiencing a burst of star formation. SAMI609396B
is analogous to a high-redshift galaxy given its low-mass and high SFR. We detect prominent, spatially resolvable emission of three auroral lines: \SIIs$\lambda\lambda$4069, 76, \OIIIs$\lambda$4363 and \SIIIs$\lambda$6312 in SAMI609396B.

In this contribution, we focus on this notable case to study direct method  metallicity and electron temperature in a spatially resolved manner.
The presence of auroral lines from multiple ionic species allows us to investigate the common assumption of using an assumed temperature relation (e.g. $T_e$(\OII) -- $T_e$(\OIII) relation) on the  spatial distribution of metallicity in galaxies.
Additionally, comparisons to strong-line metallicity trends  provide further insight into possible systematic effects in samples of gradients measured in the local and high-redshift Universe.
Given the rarity of spatially resolved $T_e$ studies at low redshift, and the relevance of this object to high-redshift comparisons, it warrants a detailed study of its own.

This work is organised as follows. In Section \ref{sec:samidata} we briefly describe the SAMI DR2 public release data, general properties of the SAMI609396 system, and selection of SAMI609396B. Our methodology for deriving spatially resolved electron temperature measurments is outlined in Section \ref{sec:te}.
In Section \ref{sec:OH_trends} we derive metallicity maps from three different ``direct method'' approaches and four different strong-line methods and discuss the differences in spatial trends favoured by each. We discuss further caveats in Section \ref{sec:discussion} before summarising and presenting conclusions in Section \ref{sec:conclusion}.
Detailed descriptions of the derivation of global properties, spectral fitting and emission line measurements are deferred to the Appendix. We also include a list of SAMI galaxies with visually identifiable auroral line emission in the Appendix.
Throughout this paper we adopt the \citet{Planck16} cosmology: $\Omega_\Lambda = 0.692$, $\Omega_M = 0.308$, $\sigma_8 = 0.815$, and $H_0 = 67.8$ km s$^{-1}$ Mpc$^{-1}$. All magnitudes are quoted in the $AB$ magnitude system \citep{Oke&Gunn83}.

\section{The SAMI Galaxy Survey}
\label{sec:samidata}

We conducted a search for auroral lines in SAMI Galaxy Survey Public Data Release 2\footnote{\url{https://sami-survey.org/abdr}} \citep{Bryant15, Green18, Scott18}.
The SAMI Galaxy Survey \citep{Bryant15} is a large IFU survey targeting low-redshift ($z\lesssim0.1$) galaxies with the Sydney -- Australian Astronomical Observatory Multi-Object Integral Field Spectrograph \citep{Croom12}. Reduced SAMI data cubes are formed by sampling dithered hexabundle observations onto a regular grid (refer to \citealt{Allen15} and \citealt{Sharp15} for details). The SAMI  aperture of radius is  approximately $\sim$7.5 arcsec with a sampling of 0$\farcs$5 $\times$ 0$\farcs$5 spaxels. The true spatial resolution is limited by the seeing, recorded as FWHM$_\textit{PSF}$ = 2.07 arcsec ($\sim$790 pc) for SAMI609396. SAMI observes in two spectral bands. The blue arm covers the observed wavelength range from 3750-5750 \AA{} at low spectral resolution (R $\sim$ 1808, $\sigma v \sim 74$ km s$^{-1}$, at 4800 \AA{}), while the red arm covers from 6300-7400 \AA{} at medium resolution (R $\sim$ 4304, $\sigma v \sim 29$ km s$^{-1}$, at 6850 \AA{}) \citep[e.g.,][]{Zhou17}.
For more detailed information on the SAMI survey and data products, the reader is referred to the above references.

Among nine SAMI galaxies in which we visually identified the presence of up to three auroral lines (\SII$\lambda\lambda$4069, 76, \OIII$\lambda$4363 and \SIII$\lambda$6312), we highlight one notable case, SAMI609396 -- a minor-merger system (Figure~\ref{fig:imaging}).
The remainder of this paper is focused on this object. The list of SAMI galaxies we compiled with identifiable auroral line emission can be found in Appendix~\ref{ap:auroral_list}.

\subsection{SAMI609396}
\label{sub:sami609396}

\begin{figure*}
	\includegraphics[width=\textwidth]{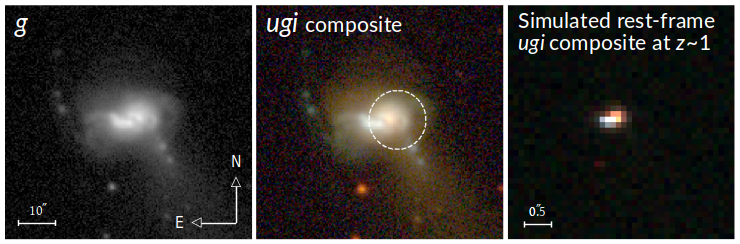}
    \caption{Left panel: $g$-band imaging of the SAMI609396 merger system from SDSS. Middle panel: $ugi$ RGB composite of the system. Prominent auroral line emission is associated with SAMI609396B, the lower-left object exhibiting strong blue colour in the $ugi$ composite. The white dashed circle in the middle panel shows the field of view observed by the SAMI IFU. The 10\farcs0 scale given for the $g$-band image applies also for the middle panel and corresponds to approximately 3.8 kpc in physical distance.
    Right panel: simulated rest-frame $ugi$ colour composite after artificially redshifting the $u$-, $g$- and $i$-band imaging to $z\sim1$. After redshifting, these bandpasses correspond approximately to \emph{HST} filters ACS/F606W, ACS/814W, and WFC3/F160W. The pixel scale in the simulated image is 0\farcs1, similar to that of \emph{HST}/WFC3. The simulated depth of the image is similar to  observations in 3D-HST \citep{Yuan20}.}
    \label{fig:imaging}
\end{figure*}

\begin{figure}
    \centering
    \includegraphics[width=\columnwidth]{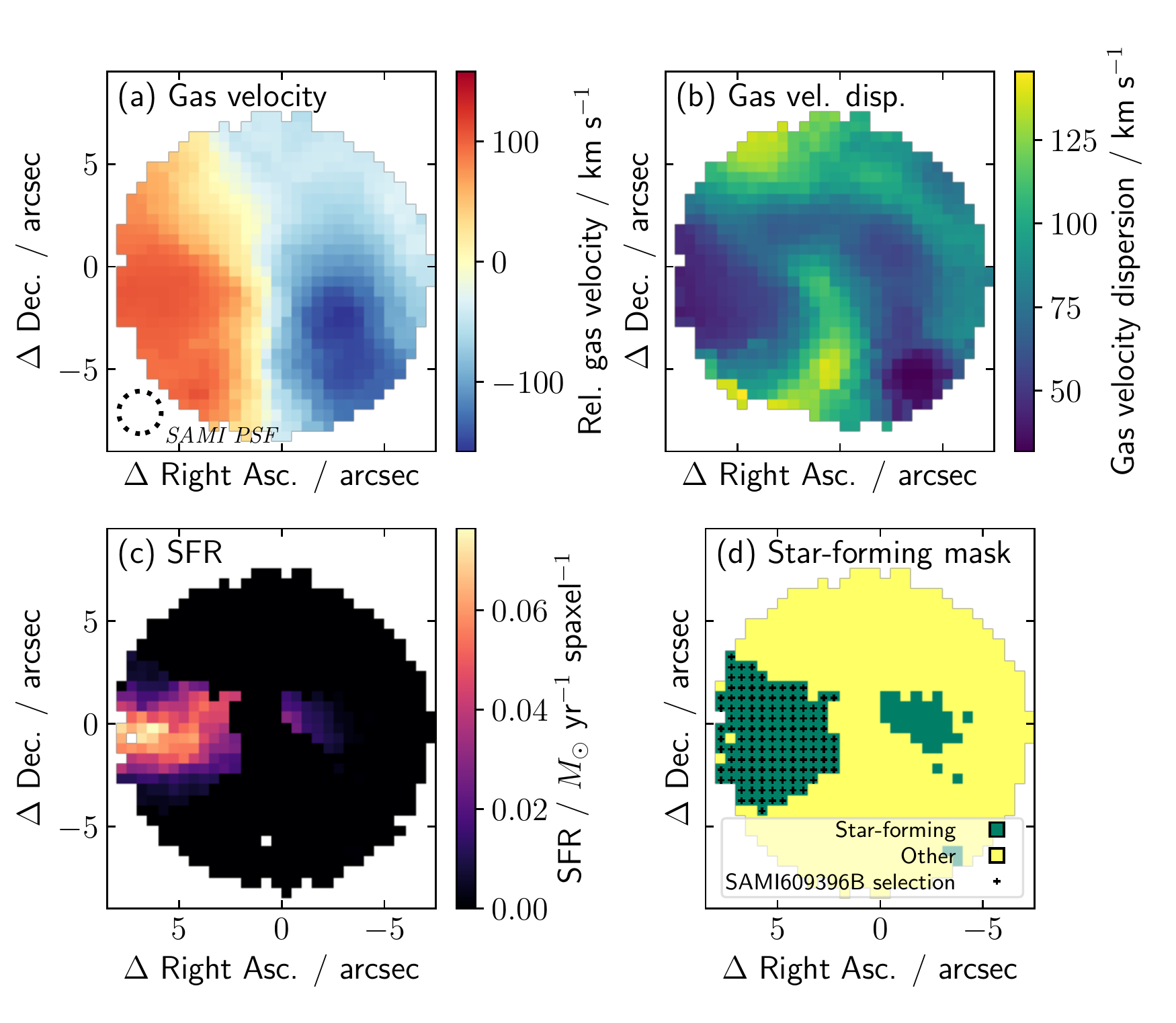}
    \caption{Publicly available value-added data products from SAMI DR2. Panel (a) Gas velocity from 1-component fitting. Panel (b) Gas velocity dispersion from 1-component fitting. Panel (c) Per spaxel star-formation rate \citep{Medling18}. Panel (d) Star-forming mask. The large star-forming dominated region denoted with black `+' symbols in panel (d) is characterised by very high SFR, velocity dispersions of $\sim$30 -- 80 km s$^{-1}$, and relative velocities of $\sim$100 km s$^{-1}$ (in the scale of panel a). This region, designated SAMI609396B, is spatially associated with observed auroral lines and is the target of our investigation. The black dotted circle in panel (a) indicates the point-spread function measured for this SAMI observation and applies to all panels.}
    \label{fig:sami_dr2}
\end{figure}

SAMI609396 (SDSS J114212.25+002004.0) is  identified as  a minor merger in the Sloan Digital Sky Survey (SDSS) images (Figure~\ref{fig:imaging}). The two merging galaxies are not deblended in the SDSS catalog with the merger system having a total $r$-band magnitude of 13.95. The SAMI input catalog gives the heliocentric redshift as $z=0.01824$.

The merger signatures are evident from the colour difference and tidal tails. A visual inspection of the system shows one smaller galaxy exhibiting a strong blue colour, with a larger companion that is significantly redder (see Figure~\ref{fig:imaging} middle panel).
Spatially-resolved 1D spectra from the publicly available SAMI datacube show that the smaller galaxy in this system (SAMI609396B) is experiencing a burst of star formation associated with strong \OIII$\lambda$5007 emission lines (Equivalent Width (EW) $\sim200$ \AA{}).
Several prominent auroral emission lines (\SII$\lambda\lambda$4069, 76, \OIII$\lambda$4363, and \SIII$\lambda$6312) are detected in SAMI609396B.
Using spatially resolved star formation rate (SFR) maps and photometry from SAMI (Appendix~\ref{ap:global_properties}), we derive SFR and $M_*$ estimates for SAMI609396B of 4.21 $\pm$ 0.30 M$_\odot$ yr$^{-1}$ and log$(M_*/M_\odot)$ {=} 9.18 $\pm$ 0.05.
These values of SFR and $M_*$ place SAMI609396B 1.3 dex above the local star formation `main-sequence' \citep{RenziniPeng15}.

\subsubsection{SAMI609396B properties in the context of high-redshift galaxies}

A number of galaxy properties have been shown to evolve systematically with redshift including SFR \citep[e.g.][]{Speagle14}, metallicity \citep[e.g.][]{Zahid13, Sanders20b}, ionisation parameter \citep{Sanders16}, and nebular emission line ratios \citep[e.g.][]{Kewley13, Steidel14}.
Given that placing observational constraints on high-redshift galaxies is comparably much more challenging than for local galaxies, there has been interest in obtaining observational constraints for ``high-redshift analogues'' \citep[e.g.][]{Heckman05, Cardamone09, Green14, Bian16}. These are galaxies at low-redshift with properties that emulate those observed in high-redshift galaxies.
Given the rarity of auroral emission lines in IFU data, we consider that SAMI609396B is worthy of a detailed study on its own. However, we also consider how its properties compare to those seen in high-redshift galaxies.

As outlined above in \S~\ref{sub:sami609396}, the SFR and $M_*$ measurements for SAMI609396B are more than 1 dex above the local star-forming main sequence, more in line with values typical of galaxies at $z\gtrsim1$.
Global metallicity correlates positively with stellar mass at $z\sim0$ (Mass-Metallicity Relation; refer to \citealt{MaiolinoMannucci19} and references therein), and at fixed stellar mass, metallicity is seen to decrease with increasing redshift \citep{Zahid13, Sanders20b}.
According to a recent multi-diagnostic determination by \citet{Sanders20b}, galaxies of a mass comparable to SAMI609396B (log$(M_*/M_\odot) \approx 9.18$) would have a median metallcity of 12+log$(O/H)=8.55$ at $z\sim0$, 12+log$(O/H)=8.26$ at $z\sim2.3$, and 12+log$(O/H)=8.17$ at $z\sim3.3$.
Absolute metallicity values for individual galaxies are notoriously difficult to determine and depend strongly on the calibration used \citep[e.g.][]{KewleyEllison08}. Although we do not take the step of applying the same metallciity calibration used by \citet{Sanders20b}, according to the metallicities we derive for SAMI609396B in \S~\ref{sec:OH_trends} we expect that the metallicity of SAMI609396B would likely fall somewhere between the median values expected from the $z\sim0$ and $z\sim2.3$ samples.

Ionisation parameters and electron densities in $z\sim2.3$ galaxies have been shown to be systematically offset from local galaxies at fixed stellar mass \citep{Sanders16}.
Electron density is most commonly probed with the \SIIs$\lambda$6716/$\lambda$6731 doublet ratio. MOSDEF galaxies at $z\sim2.3$ were found by \citet{Sanders16} to have a median \SIIs doublet ratio of 1.13, corresponding densities of around 290 cm$^{-3}$ in the S$^+$ zone of emitting nebulae, much higher than typical SDSS values (\SIIs$\lambda$6716/$\lambda$6731 = 1.41; density of 26 cm$^{-3}$).
We measure a global \SIIs ratio of 1.29 for SAMI609396B, corresponding to a density of 118 cm$^{-3}$, placing SAMI609396B between the low- and high-redshift sample medians. Given the scatter about those median values in both the MOSDEF and SDSS samples (Figures 4 \& 5 in \citealt{Sanders16}), it is difficult to draw conclusions about how SAMI609396B compares to the two populations based on density.
Using the $O_{32}$\footnote{$O_{32}$ = \OIIIs $\lambda\lambda$4959, 5007 / \OIIIs $\lambda\lambda$3726, 29 in this context} strong-line ratio as a tracer for ionisation parameter, \citet{Sanders16} found that, like SDSS galaxies, $z\sim2.3$ MOSDEF galaxies show a trend of decreasing ionisation parameter with increasing stellar mass. They find the slope of this relation to be very similar to that of SDSS galaxies, however with a $\sim$0.6 dex offset toward higher O32 at fixed stellar mass in the $z\sim2.3$ sample (Fig 8 in \citealt{Sanders16}).
Given the stellar mass derived for SAMI609396B (log$(M_*/M_\odot) \approx 9.18$), SDSS galaxies have a median value of $O_{32} = -0.25$, while MOSDEF galaxies with comparable mass have much higher values ($O_{32} = 0.28$). We measure $O_{32} = 0.14$ for SAMI609396B, 0.39 dex higher than the median SDSS value and 0.14 dex below the median of $z\sim2.3$ MOSDEF galaxies.

Recent studies have found an offset in the locus inhabited by high-redshift galxies on the $N2$-BPT diagram \citep{BPT81} with high-redshift galaxies exhibiting higher \OIII/H$\beta$ at fixed \NII/\Has \citep{Kewley13, Steidel14}. We find that BPT line ratios observed for SAMI609396B to be within the range  of local galaxies. Our analysis of the spatially resolved BPT diagram of SAMI609396B is discussed in detail in \S~\ref{sec:discussion}.


To summarise, we find that the physical properties (SFR and ISM conditions) of SAMI609396B tend to be offset from median $z\sim0$ values, although are generally less extreme than $z\sim2$ galaxies.
In combination with the high EW(\OIII), we consider that the physical properties of SAMI609396B might be analogous to intermediate-redshift ($0 < z \lesssim 1$) galaxies.
Low-mass galaxies like SAMI609396B are extremely difficult to resolve  at high redshift.
To visually demonstrate what a system like SAMI609396 would look like at a higher redshift, we simulate the angular size and morphology of SAMI609396 at $z\sim1$ using similar techniques to those detailed in \citet{Yuan20}. The redshifted morphology is presented on the right panel of Figure~\ref{fig:imaging}. In order to resolve a low-mass system like SAMI609396B at $z\sim1$ with comparable physical resolution of SAMI,  a minimal angular resolution of 0.1" is required. Such a fine resolution can be achieved either through ground-based adaptive optics or space instruments.  The faintness of these low-mass systems also means the need for
next-generation facilities such as JWST/NIRSpec and ground-based ELTs.

\subsection{SAMI DR2: Value-added data products}
\label{sub:value_added}

SAMI DR2 includes a number of publicly available value-added data products, which we use to guide our initial understanding of the SAMI609396 system. Figure~\ref{fig:sami_dr2} shows publicly available maps for the gas velocity, gas velocity dispersion, and star-formation rate (Panels (a) -- (c)) derived from 1-component fits.

Panel (d) of Fig~\ref{fig:sami_dr2} shows a star-formation mask, determined according to \citealt{Kewley06} based on BPT \& VO87 diagnostic diagrams \citep{BPT81, VO87}, with green denoting spaxels passing selection as ``star-formation dominated''. Figure~\ref{fig:sami_dr2} shows that much of the SAMI field-of-view is dominated by emission from non-star-forming sources (yellow spaxels; ``other''). The yellow spaxels have  higher velocity dispersion compared with star-forming dominated regions, characteristic of emission from shock-heated gas. The BPT diagram and the origin of emissions in these regions are discussed further in \S \ref{sub:bpt}.

The prominent auroral line emission we identify is spatially associated with the large star-formation dominated region in the left-hand (eastern) portion of the star-formation mask. This region has a median rest-frame gas velocity of $v_\textit{gas} \approx 100$ km s$^{-1}$ (refer to scale in Fig~\ref{fig:sami_dr2}), a velocity dispersion of range $\sigma_\textit{gas} \approx 30 - 80$  km s$^{-1}$, and high a star-formation rate (median SFR surface density $\approx 0.97$ $M_\odot$ yr$^{-1}$ kpc$^{-2}$).

We designate this object as ``SAMI609396B'' and define its selection within the SAMI609396 datacube as including spaxels labelled as star-formation dominated with $v_\textit{gas} > 0$, denoted by black `+' symbols in panel (d) of Fig~\ref{fig:sami_dr2}. Global SFR and stellar mass estimates for SAMI609396B and its companion galaxy are provided in Table~\ref{tab:sami_properties}. Details of how these are derived are provided in Appendix~\ref{ap:global_properties}.

\begin{table}
    \caption{Global properties of SAMI609396B and its companion.}
    \label{tab:sami_properties}
    \begin{tabular}{ll}
        \hline
        \hline
        Right Ascension~~~~~~~~~~~ & 11$^{\rm h}$ 42$^{\rm m}$ 12\fs25 \\
        Declination & +00\degr 20$^{\rm m}$ 04\fs04 \\
        $z$ & 0.01824 \\
        \hline
        \hline
        \multicolumn{2}{l}{SAMI609396B:}\\
        \hline
        SFR (M$_\odot$ yr$^{-1}$) $^\text{a}$ & 4.21 $\pm$ 0.30 \\
        log$(M_*/M_\odot)$ & 9.18 $\pm$ 0.05 \\
        \hline
        \hline
        \multicolumn{2}{l}{Companion:}\\
        \hline
        SFR (M$_\odot$ yr$^{-1}$) $^\text{a}$ & 0.32 $\pm$ 0.08 \\
        log$(M_*/M_\odot)$ & 9.88 $\pm$ 0.07 \\
        \hline
        \hline
        \multicolumn{2}{l}{$^\text{a}$SFR measurement for area within SAMI FoV (see Fig~\ref{fig:imaging}).}\\
        \multicolumn{2}{l}{This is best considered as a lower bound.}
    \end{tabular}
\end{table}




\section{Spatially Resolved Electron Temperature}
\label{sec:te}

The electron temperature ($T_e$) and electron density ($n_e$) are fundamental physical parameters in understanding the emission line physics of ionized nebulae. Abundance measurements from collisionally excited lines in \HII regions are very sensitive to these parameters. For this reason, chemical abundances derived following explicit measurements of $T_e$ and $n_e$ are generally used as a baseline calibration for understanding the chemistry of ionized nebulae \citep[e.g.][]{MaiolinoMannucci19}.

This is generally achieved with the so-called ``direct method'' via measurement of an auroral emission line and a strong nebular line of the same ionic species. This is most commonly applied to the O$^{2+}$ ion using the \OIIIs$\lambda$4363/$\lambda$5007 ratio, which is primarily sensitive to $T_e$ (its $n_e$ dependence is minimal over the density range of typical \HII regions). Within the typical rest-frame near-ultraviolet to near-infrared wavelength range observed for galaxies, auroral line ratios may be observable for a number of ionic species including
O$^{+}$, N$^{+}$, S$^{2+}$ and S$^{+}$,
each of which probe different zones within the emitting \HII regions according to the distribution of those ions within the nebular structure. Although we detect auroral lines from three ionic species in SAMI609396B (\SII, \OIIIs and \SIII), we are able to derive electron temperature for only the \OIIIs and \SIIs ionisation zones as we lack the spectral coverage to measure the \SIIIs$\lambda$9069 and \SIIIs$\lambda$9531 strong lines required to derive $T_e$(\SIII).

\subsection{Auroral Emission Line Measurements}
\label{sub:auroral}

We derive flux maps for auroral lines from three ionic species (\SIIs$\lambda\lambda$4069, 76, \OIIIs$\lambda$4363, and \SIIIs$\lambda$6312) identified in the SAMI609396 data cube, as the SAMI DR2 value-added data products do not include emission line maps for these fainter lines. We concomitantly re-derive strong emission line fluxes, rather than use SAMI DR2 emission line maps, ensuring self-consistency in our line ratio measurements. These flux maps are generated by applying standard methods to each spaxel, first fitting the stellar continuum, and then simultaneously fitting profiles to each emission line included in our analysis. Details of this spectral fitting are provided in Appendix~\ref{ap:fitting}.

We obtain $S/N \sim 3-15$ in individual spaxels for each of \OIIIs$\lambda$4363, \SIIs$\lambda\lambda$4069, 76 and \SIIIs$\lambda$6312 across the majority of the spatial region selected as SAMI609396B.
We identify from visual inspection some degree of blending between \OIIIs$\lambda$4363 and a neighbouring faint \FeII emission line at $\lambda$4360, similar to that observed in other recent studies \citep[e.g.][]{Curti17, Berg20, ArellanoCordova20}.
We find that the \OIIIs$\lambda$4363 line is brighter than the $\lambda$4360 feature by a factor of $\sim$2 and that with the spectral resolution of the blue arm of the SAMI spectrograph we are able to reliably recover the  \OIIIs$\lambda$4363 flux. Our efforts to test the reliability of our \OIIIs$\lambda$4363 flux measurements are outlined in detail in Appendix \ref{apsub:FeOIIIblending}.

\begin{figure}
    \centering
    \includegraphics[width=\columnwidth]{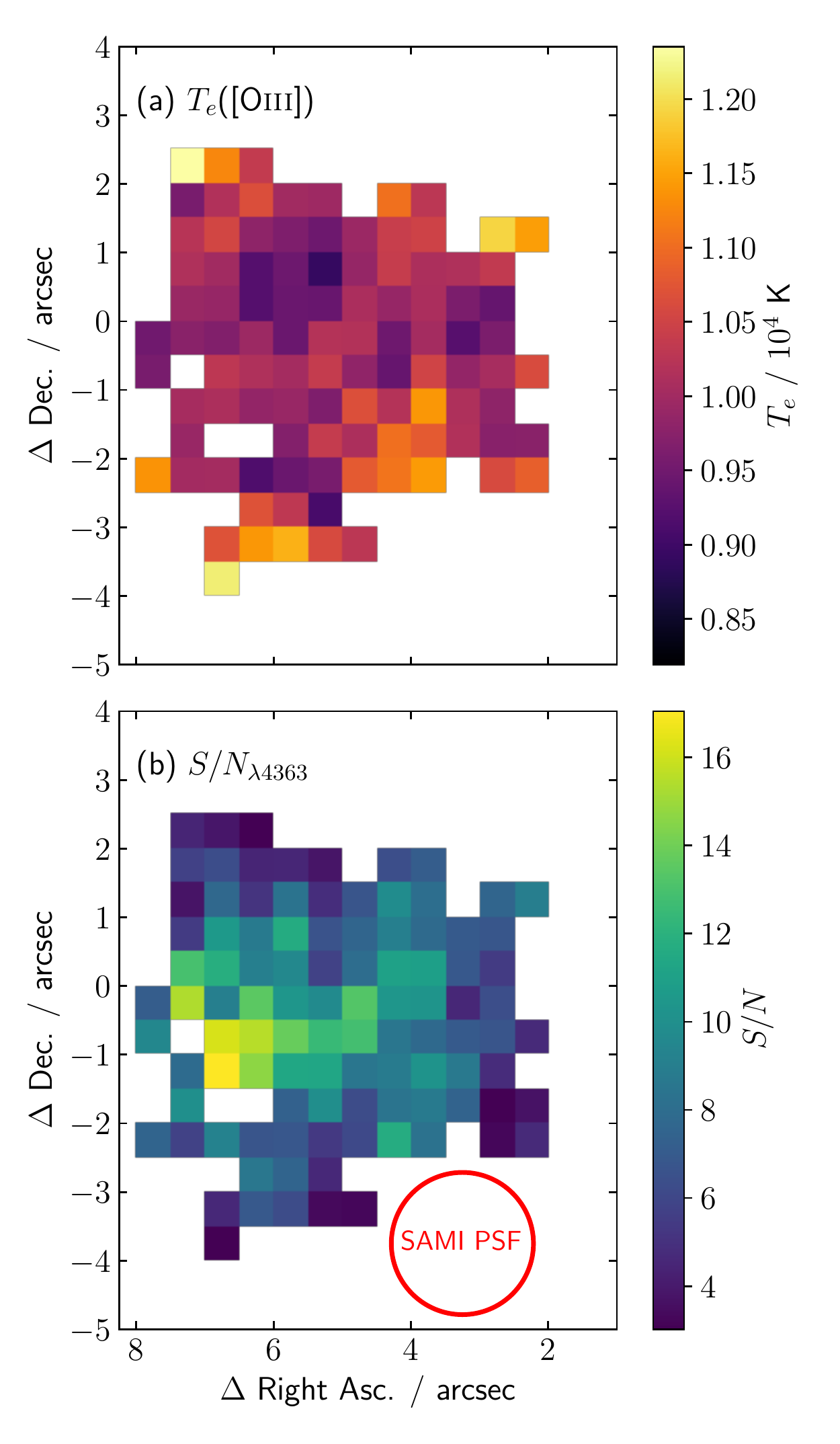}
    \caption{\textit{Top}: Electron temperature map derived from the \OIII$\lambda$4363 / \OIII$\lambda$5007 ratio (see \S \ref{sub:te_OIII}). \textit{Bottom}: Measured signal-to-noise of \OIII$\lambda$4363 auroral line. The red circle depicts the FWHM PSF of this SAMI datacube and applies to both panels. The large scale spatial variations in $T_e$ do not appear to correlate with \OIII$\lambda$4363 $S/N$.}
    \label{fig:tOIII_map}
\end{figure}

\subsection{[OIII] Electron Temperature}
\label{sub:te_OIII}

The emission line ratio most widely used to determine the electron temperature with the direct method is the \OIIIs$\lambda$4363 / \OIIIs$\lambda$5007 ratio. Despite the primary dependence of this \OIIIs ratio on temperature, the residual density dependence is often accounted for by measurement of a density sensitive line ratio, typically \SIIs$\lambda$6716 / \SIIs$\lambda$6731. \citet{Izotov06} use relations derived for these aforementioned \OIIIs and \SIIs line ratios (Equations (1) and (2) in that reference) in an iterative manner, solving simultaneously for $T_e$ and $n_e$. This iterative approach is shared by the \texttt{getCrossTemDen} routine in the \texttt{PyNeb} package \citep{PyNeb15}, which allows for a flexible array of temperature- and density-sensitive line ratios.

However, it is important to consider that neither temperature nor density is expected to be constant throughout \HII regions. Additionally, emission from different ionic species may not be co-spatial. Certainly, \SIIs emission is expected to arise from the outer regions of nebulae, thus densities measured from the \SIIs line ratio do not necessarily provide a good indication of the density of the \OIIIs emission region (see Figure 2 in \citealt{Kewley19}).

Given these uncertainties, \citet{Nicholls20} instead propose a simplified approach in which $T_e$ is derived from an empirical relation of the auroral line ratio, derived from \HII region modelling, forgoing any attempt to account for $n_e$, suggesting that any improvements in temperature insight are outweighed by uncertainties induced by density variations and lack of co-spatiality.

Given the $\sim$1 kpc spatial resolution of SAMI, we are unable to resolve individual \HII regions, adding to the uncertainties described above.
Thus, we use this simplified approach to derive our $T_e$ from the \OIIIs$\lambda$4363 / \OIIIs$\lambda$5007 ratio according to the relation given in \citet{Nicholls20}. This relation is shown as Equation~\ref{eq:temp_OIII} here:

\begin{equation}
\label{eq:temp_OIII}
    \text{log}_{10}(T_e(\text{[OIII]})) = \frac{3.3027 + 9.1917x}{1.0 + 2.092x - 0.1503x^2 - 0.0093x^3}
\end{equation}{}

where $x = \text{log}_{10}(f_{4363}/f_{5007})$, with $f_X$ referring to a line flux measurement of a collisionally excited line with rest-frame wavelength X \AA{}, and $T_e$ is in units of K.
The derived \OIIIs temperature map for spaxels with \OIIIs$\lambda$4363 of S/N $>3$ is shown in Figure~\ref{fig:tOIII_map}.

\subsection{[SII] Electron Temperature}
\label{sub:te_SII}

\begin{figure*}
    \centering
    \includegraphics[width=\textwidth]{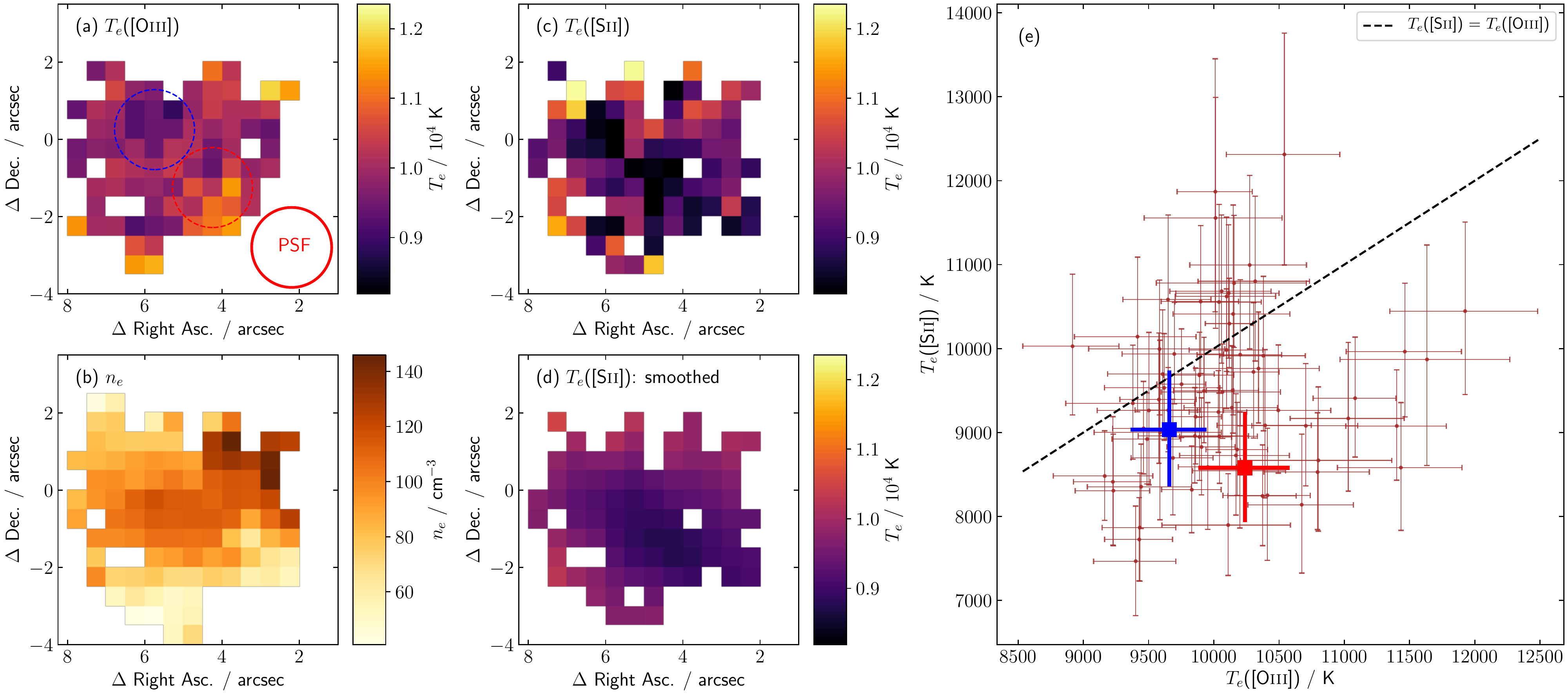}
    \caption{Comparison of $T_e$(\OIII) and $T_e$(\SII) electron temperature values. \textit{Panel (a)}: map of $T_e$(\OIII) values for spaxels with $S/N > 5$ for \OIIIs$\lambda$4363. Red circle labelled `PSF' has diameter equal to the FWHM of the SAMI PSF for this observation and applies to panels (a--d). \textit{Panel (b)}: electron density derived from the \SIIs$\lambda$6716 / $\lambda$6731 ratio. \textit{Panel (c)}: map of $T_e$(\SII) values for spaxels with $S/N > 5$ for \SIIs$\lambda\lambda$4069, 76. \textit{Panel (d)}: map of $T_e$(\SII) from panel (c) smoothed with a Gaussian filter. \textit{Panel (e)}: Brown points show values of of $T_e$(\SII) and $T_e$(\OIII) for individual spaxels with $S/N > 5$ on both auroral lines. Error bars shown reflect only measurement uncertainty and do not include associated modelling uncertainties. Temperatures derived for two mock apertures (indicated by blue and red dashed circles in panel a) are shown as the blue and red points in panel (e).}
    \label{fig:tSII_v_tOIII}
\end{figure*}

In addition to $T_e$(\OIII), spatially resolved measurements of the \SIIs auroral lines allow us to measure $T_e$(\SII) from the \SIIs$\lambda\lambda$4069, 76 / \SIIs$\lambda\lambda$6716, 31 ratio.

Modelling indicates that at the low density limit ($1<n_e<50$ cm$^{-3}$), the residual density dependence of the \SIIs$\lambda\lambda$4069, 76 / \SIIs$\lambda\lambda$6716, 31 ratio is minimal. In contrast to the \OIIIs case, this \SIIs temperature diagnostic is co-spatial with the \SIIs density diagnostic, meaning that we are able to make a more reliable estimate of the density.
The $n_e$ values for SAMI609396 obtained with the \SIIs$\lambda$6716 / \SIIs$\lambda$6731 ratio \citep[Eq. 3 in ][]{Proxauf14} are shown in Fig~\ref{fig:tSII_v_tOIII} Panel (b). We find the median electron density to be $\widetilde{n_e} = 92$ cm$^{-3}$. This value is above the \SIIs\ low density limit, indicating that \SIIs$\lambda\lambda$4069, 76 / \SIIs$\lambda\lambda$6716, 31 will have a residual density dependence.
Nonetheless, we derive the \SIIs\ temperature with a similar approach to that outlined in \S \ref{sub:te_OIII} with a new rational polynomial fit to modelling data assuming a density of $n_e=100$ cm$^{-3}$. This fit is given in Equation~\ref{eq:temp_SII} where $x = \text{log}_{10}[(f_{4069}+f_{4076})/(f_{6716}+f_{6731})]$ and $T_e$ is in units of K.

\begin{equation}
\label{eq:temp_SII}
    \text{log}_{10}(T_e(\text{[SII]})) = \frac{-0.08891 + 2.06354x + 3.38680x^2 + 0.10754x^3}{0.1 + 0.78000x + 0.94404x^2}
\end{equation}{}

$T_e$(\SII) values obtained for SAMI609396 are compared with $T_e$(\OIII) values in Figure~\ref{fig:tSII_v_tOIII}. Panels (a) and (b) show maps of $T_e$(\SII) and $T_e$(\OIII) respectively for spaxels where the relevant auroral line is detected with $S/N>5$. Panel (e) shows the direct comparison of $T_e$(\SII)) and $T_e$(\OIII) values on a spaxel-by-spaxel basis.
We observe that a majority of points in panel (e) of Fig~\ref{fig:tSII_v_tOIII} lie below the line of $T_e$(\SII) = $T_e$(\OIII) (i.e. higher $T_e$(\OIII) than $T_e$(\SII)).
The large blue and red points in Fig~\ref{fig:tSII_v_tOIII} Panel (e) show derived $T_e$(\SII) and $T_e$(\OIII) electron temperatures for two mock apertures which correspond to the regions shown as blue and red dashed circles in Panel (a).
These aperture temperatures appear to indicate that $T_e$(\SII) and $T_e$(\OIII) do not exhibit strong positive correlation across different spatial regions of SAMI609396B. The implications of this temperature relation for metallicity measurement are discussed further in \S \ref{sec:OH_trends}.




\section{Spatial Trends In Metallicity}
\label{sec:OH_trends}

In Section~\ref{sec:te} we derived spatially resolved electron temperature ($T_e$) measurements.
Here we use these $T_e$ measurements to determine direct method oxygen abundances under three different sets of assumptions, showing that derived spatial variations in metallicity can be very sensitive to the assumed internal \HII region temperature structure.
Additionally we derive spatially resolved strong-line metallicities and discuss differences in observed spatial trends.

\subsection{Direct Method Metallicity}
\label{sub:direct_logOH}

\begin{figure*}
    \centering
    \includegraphics[width=\textwidth]{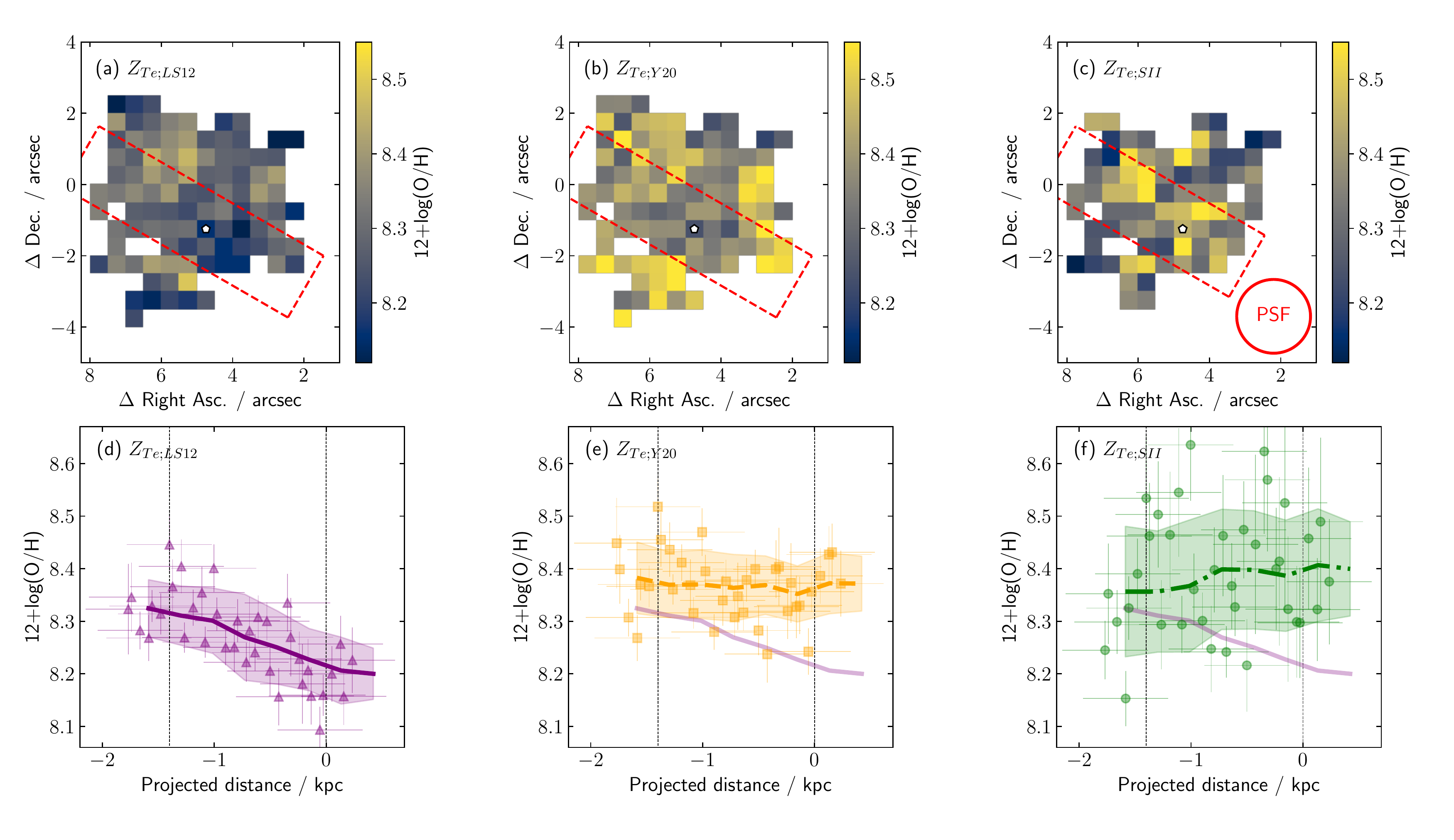}
    \caption{
    Observed spatial trends in direct method metallicity depend strongly on temperature structure assumptions.
    Direct method metallicity maps (panels a-c) and spatial metallicity trends (d-f) are shown for SAMI609396B under three different $T_e$(\OII) temperature assumptions.
    Panels (a, d) show \ZLS: where $T_e$(\OII) is derived from $T_e$(\OIII) via the relation of \citet{LopezSanchez12} (Eq~\ref{eq:tOII_tOIII}).
    Panels (b, e) show \ZYates: derived as for \ZLS with the additional step of applying the empirical correction of \citetalias{Yates20} based on $O32$.
    Panels (c, f) show \ZSII: metallicity is derived assuming $T_e$(\OII) = $T_e$(\SII).
    See \S \ref{sub:direct_logOH} for details.
    Maps in panels (a-b) include spaxels with $S/N_{\lambda4363} \geq 3$, while panel (c) additionally excludes spaxels with $S/N_{\lambda4069} < 3$. The red circle in panel (c) shows the FWHM of the SAMI PSF and applies to panels (a-c). The dashed red rectangles in panels (a-c) span the region of highest S/N for the \OIII$\lambda$4363 line and defines the spatial region examined in panels (d-f).
    Panels (d-f) show individual points for which $S/N_{\lambda4363} > 5$. Trend lines indicate running medians of the points shown. Vertical error bars on individual points reflect only measurement uncertainties and are dominated by auroral line measurements.
    }
    \label{fig:direct_metal_map}
\end{figure*}

Since the abundance of neutral oxygen (O$^0$) and oxygen in ionization states higher than O$^{2+}$ is expected to be negligible in \HII regions, we assume that the total oxygen abundance can be approximated as Equation~\ref{eq:oxygen_abundance}:

\begin{equation}
\label{eq:oxygen_abundance}
    \frac{O}{H} = \frac{O^+}{H^+} + \frac{O^{2+}}{H^+} .
\end{equation}

\smallskip

We derive abundances of these two ionisation states of oxygen using the following analytic relations set out in \citet{PerezMontero17}:

\begin{multline}
\label{eq:PM17_O3_abundance}
    12+\text{log}\left(\frac{O^{2+}}{H^+}\right) = \text{log}\left(\frac{f_{4959}+f_{5007}}{f_{H\beta}}\right) + 6.1868 \\ + \frac{1.2491}{t(O^{2+})} - 0.5816\cdot\text{log}\Big(t(O^{2+})\Big)
\end{multline}

\begin{multline}
\label{eq:PM17_O2_abundance}
    12+\text{log}\left(\frac{O^{+}}{H^+}\right) = \text{log}\left(\frac{f_{3726}+f_{3729}}{f_{H\beta}}\right) + 5.887 + \frac{1.641}{t(O^{+})} \\ - 0.543\cdot\text{log}\Big(t(O^{+})\Big) + 0.000114\cdot n_e
\end{multline}

where $t(O^{2+}) = T_e$(\OIII)$/10^4$ K, $t(O^{+}) = T_e$(\OII)$/10^4$ K, $n_e$ is the electron density measured by the \SIIs $\lambda$6716 / $\lambda$6731 ratio, and $f_X$ refers to a line flux measurement of the H$\beta$ Balmer line or a collisionally excited line with rest-frame wavelength X \AA{}. Deriving O$^{2+}$/H$^+$ in this way requires only \OIIIs$\lambda\lambda$4959, 5007 and H$\beta$ emission line fluxes in addition to the $T_e$(\OIII) values derived in \S~\ref{sub:te_OIII}. On the other hand, the O$^+$/H$^+$ abundance from Eq.~\ref{eq:PM17_O2_abundance} calls for $T_e$(\OII), which we do not directly measure. Additionally, O$^+$/H$^+$ has residual dependence on $n_e$, although we simply adopt the same fixed denisty $n_e = 100$ cm$^{-3}$ used in the temperature calculations in \S \ref{sub:te_OIII}. Note that our derived metallicity values vary by less than 0.01 dex with changes in adopted density, provided those are below $n_e < 200$ cm$^{-3}$.

Unlike $T_e$(\OIII), we do not directly measure $T_e$(\OII), since we are unable to detect either the \OIIs$\lambda\lambda$7319, 30 or \OIIs$\lambda\lambda$2470+ doublets.
A favourable alternative is to use temperatures derived from other ionic species, especially \NIIs or \SIIIs, to probe the temperature structure \citep[e.g.][]{Berg20}.
However, given the faintness of auroral lines it is common that an observation may enable measurement of only the \OIIIs temperature zone.
In this scenario, a $T_e$(\OII) estimate can be obtained by adopting an empirical $T_e$(\OII) -- $T_e$(\OIII) relation, for which a number of calibrations exist \citep[e.g.][]{Izotov06, LopezSanchez12}.
Despite expanding the number of observations for which direct metallicities can be derived, \citet{Yates20} (\citetalias{Yates20} hereafter) find that using $T_e$(\OII) -- $T_e$(\OIII) relations can underestimate the direct metallicity by more than 0.5 dex for low-ionisation systems, highlighting the importance of constraining the internal temperature structure of \HII regions where possible. Additionally, \citetalias{Yates20} provide an empirical correction for this effect based on the \OIII/\OIIs strong line ratio.

For this analysis, we determine our total oxygen abundance maps in three ways. Each differs in its approach to handling the O$^{+}$/H$^+$ abundance, while in all three cases the O$^{2+}$/H$^+$ abundance is determined from Eq~\ref{eq:PM17_O3_abundance} and our direct measurement of $T_e$(\OIII). For the remainder of this paper, metallicities derived in these three ways will be abbreviated as $Z_\text{Te; LS12}$, $Z_\text{Te; Y20}$ and $Z_\text{Te; SII}$ (where $Z=12+\text{log}(O/H)$), described as follows:
\begin{enumerate}
    \item $Z_\text{Te; LS12}$: O$^{+}$/H$^+$ is determined using $T_e$(\OII) derived from  $T_e$(\OIII) using the relation outlined in \citet{LopezSanchez12} (Eq~\ref{eq:tOII_tOIII}).\footnote{We note that alternative $T_e$([OII]) -- $T_e$([OIII]) relations, including the equations from \citet{Izotov06}, do not significantly affect the metallicity morphology obtained for SAMI609396B.} This is the most commonly adopted method.
    \smallskip
    \item $Z_\text{Te; Y20}$: As for $Z_\text{Te; LS12}$, with the subsequent application of the \citetalias{Yates20} empirical correction, based on \OIII/\OIIs strong-line ratio (Eq~\ref{eq:y20_correction}). This is a relatively new correction and has not been widely implemented in literature yet.
    \smallskip
    \item $Z_\text{Te; SII}$: O$^{+}$/H$^+$ is determined with $T_e$(\OII) derived instead from $T_e$(\SII) using the assumption $T_e$(\OII) = $T_e$(\SII). This is uniquely enabled by the detection of \SII\ auroral lines in this study.
\end{enumerate}

\subsubsection{Empirical $T_e$(\OII) -- $T_e$(\OIII) relation}

For \ZLS we adopt the $T_e$(\OII) -- $T_e$(\OIII) relation as calibrated by \citet{LopezSanchez12}, given in Equation \ref{eq:tOII_tOIII}:

\begin{equation}
\label{eq:tOII_tOIII}
    T_e\text{\OII} = T_e\text{\OIII} + 450 - 70 \cdot \text{exp}\big[(T_e\text{\OIII}/5000)^{1.22}\big]
\end{equation}

Deriving $T_e$(\OII) in this way and applying Equations~\ref{eq:PM17_O3_abundance}--\ref{eq:PM17_O2_abundance} we obtain the total oxygen abundance map shown in panel (a) of Fig~\ref{fig:direct_metal_map}. The spatial structure of this map reflects that of the temperature map derived in Fig~\ref{fig:tOIII_map} and favours a strong trend in metallicity across the region of the highest signal-to-noise (Fig~\ref{fig:direct_metal_map}, panel d).

The measurement uncertainty is dominated by the flux uncertainty of the \OIIIs$\lambda$4363 emission line to the point where the measurement uncertainty contribution from the high $S/N$  \OIII, \OIIs and H$\beta$ strong lines can be ignored. We see no obvious correlation between the $S/N$ of \OIIIs$\lambda$4363 and $T_e$(\OIII) (Fig~\ref{fig:tOIII_map}). Increasing the minimum $S/N$ cut on the \OIIIs$\lambda$4363 auroral line from $S/N>3$ to $S/N>8$ changes the median metallicity by less than 0.005 dex. Together, these give us confidence that observed spatial variations in metallicity are not artifacts from measurement noise, although the effects of modelling uncertainty are discussed over the coming sections.

\subsubsection{Empirical O$^{+}$ Abundance Correction}
\label{sub:y20_correction}

\citet{Yates20} provide an empirical correction based on the observed \OIII/\OIIs line ratio given by Equation \ref{eq:y20_correction},

\begin{equation}
\label{eq:y20_correction}
    Z_\text{Te; Y20} = Z_\text{Te; LS12} - 0.71 \cdot (O32 - 0.29)
\end{equation}

where $Z_\text{Te; Y20}$ and $Z_\text{Te; LS12}$ are corrected and uncorrected values of 12+log(O/H) respectively; $O32$ = log(\OIIIs$\lambda\lambda$4959, 5007 / \OIIs$\lambda\lambda$3726, 9) and the correction is applied only when $O32 \leq 0.29$.

Values of $O32$ across SAMI609396B fall in the range for which this correction will be non-zero.
Our direct metallicity map after \citetalias{Yates20} correction is shown in Fig~\ref{fig:direct_metal_map} Panel \textit{(b)}.
Spatial variations in the $O32$ ratio result in a flattening of the spatial trend after application of this correction.


We note that, in addition to the empirical correction described here (``\citetalias{Yates20} correction''), \citet{Yates20} also outlined a novel method for determining semi-direct metallicities (``\citetalias{Yates20} method'') in which $T_e$(\OII) and metallicity are solved for simultaneously, rather than sequentially. This \citetalias{Yates20} method then also requires subsequent application of the \citetalias{Yates20} correction if $O32 \leq 0.29$, as above. Note that Figure 6 in \citet{Yates20} shows that the abundance deficit at low $O^{++}/O^+$, which the \citetalias{Yates20} correction adjusts for, is present to varying degrees for all $T_e$(\OII) -- $T_e$(\OIII) relations considered in that work.

We find the \citetalias{Yates20} method  gives a two-valued solution for SAMI609396B which may require an additional prior to select the best metallicity solution.
We found that applying the \citetalias{Yates20} method as originally outlined favoured the lower value of these two solutions which yielded a gradient comparable to that obtained from our \ZYates approach here, albeit with a much lower normalisation ($\sim$0.3 dex).
We found that the normalisation of the upper-branch solution was in better agreement with our other determinations outlined here, however the spatial trend arising from this upper-branch solution is more difficult to interpret.
Discussion of our implementation of the \citet{Yates20} method and its two-valued nature is deferred to Appendix~\ref{ap:yates_method}.


\subsubsection{O$^{+}$ abundance with $T_e$(\SII)}
\label{sub:ZSII}

The \SIIs temperature samples a relatively narrow zone from the outer regions of nebulae and is consequently not widely used to constrain the temperature profile of emitting \HII regions. However, \citet{Croxall16} found general agreement of $T_e$(\SII) with $T_e$(\OII) and $T_e$(\NII) in \HII regions in NGC 5457. In the absence of the \SIIIs strong-lines, the \NIIs auroral lines, or any other temperature probes, $T_e$(\SII) affords our only direct probe of the internal temperature structure of \HII regions in SAMI609396B.

We make the simplified assumption that $T_e$(\OII) = $T_e$(\SII) and update our total oxygen abundance using the measured $T_e$(\SII) map (Fig~\ref{fig:tSII_v_tOIII} panel c) to re-derive our O$^+$/H$^+$ values. These updated oxygen abundances are shown in Fig~\ref{fig:direct_metal_map} Panel (c), spanning a slightly smaller spatial extent due to the additional requirement of \SIIs auroral line signal-to-noise. The spatial trend shown in Fig~\ref{fig:direct_metal_map} Panel (f) is seen to be opposite of that in Panel (d) where O$^+$/H$^+$ was derived using an empirical temperature relation, albeit with a larger scatter.

This stark reversal can be explained by the $T_e$(\SII) -- $T_e$(\OIII) trends observed in Figure~\ref{fig:tSII_v_tOIII}.
Deriving $T_e$(\OII) from a relation with $T_e$(\OIII) assumes that such a relation is fixed across the spatial region covered. This would mean that regions with elevated $T_e$(\OIII) would also show increased $T_e$(\OII).
However, the apertures plotted in panel (e) of Fig~\ref{fig:tSII_v_tOIII} (blue and red bold points) show that despite the increase in $T_e$(\OIII) from the `blue' aperture to the `red' aperture, measured $T_e$(\SII) instead decreases (albeit with large uncertainties). This suggests the absence of a strong positive correlation between these temperatures across the spatial region and highlights the limitations of applying empirical temperature relations to measure spatial metallicity trends.
This is likely driven by variations in the ionisation structure (i.e. $O^{2+}/O^+$ abundance ratio) and also explains the observed variations in $O32$ ratio that lead to the flattening of the spatial trend observed after applying the \citetalias{Yates20} correction.
We discuss this further in \S~\ref{sub:gradient}.

\begin{figure*}
    \centering
    \includegraphics[width=\textwidth]{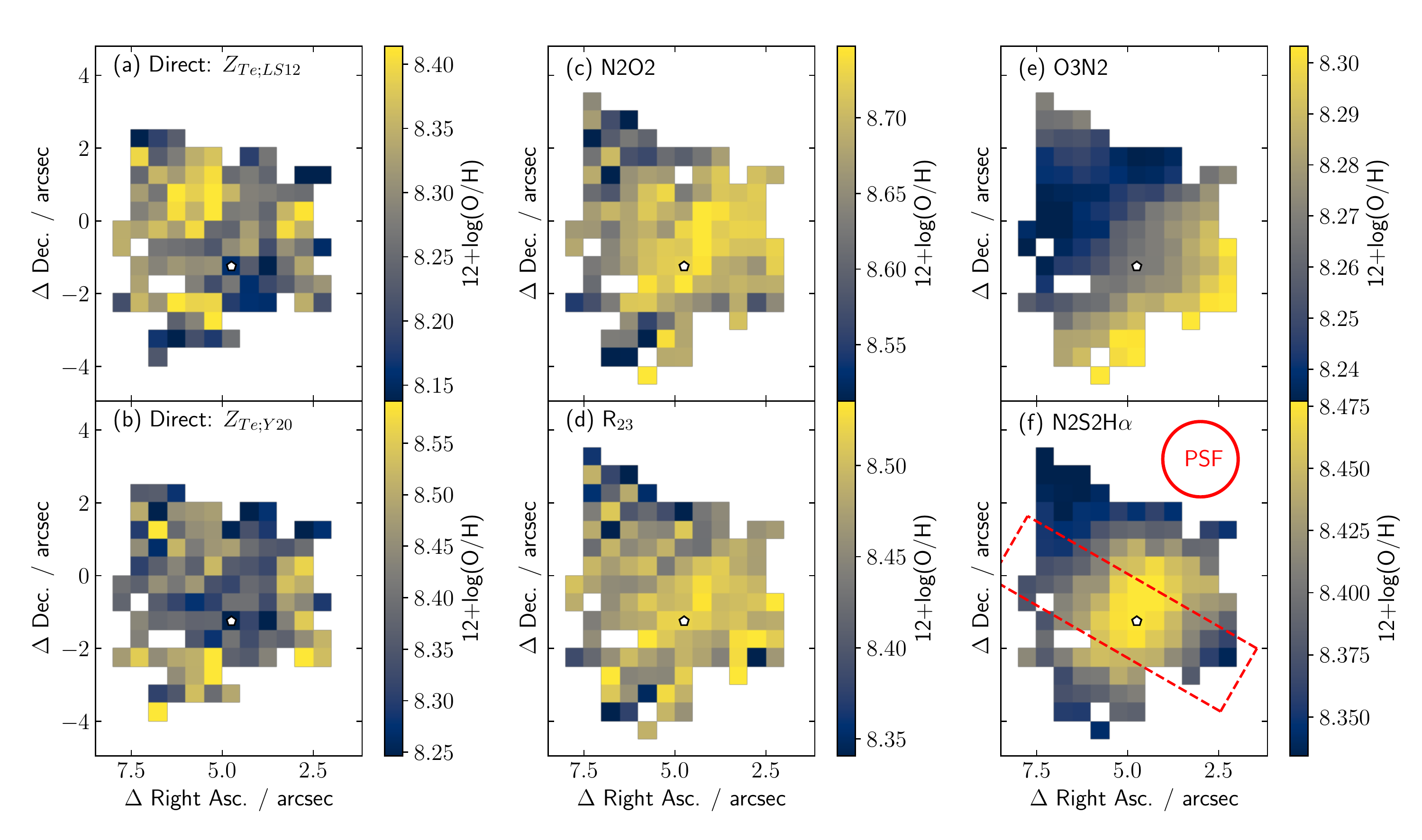}
    \caption{Direct method and strong-line oxygen abundance maps for the star-formation selected region corresponding to SAMI609396B. \textit{Panel (a)}: Direct method metallicity using $T_e$ values derived from \OIIIs$\lambda$4363 / $\lambda$5007 ratio (see \S \ref{sub:direct_logOH}).
    \textit{Panel (b)}: Direct method metallicity after applying the empirical correction of \citet{Yates20} (see \S \ref{sub:y20_correction}).
    \textit{Panel (c)}: iterative solution for metallicity, solved simultaneously for metallicity with $N2O2$ and ionisation parameter with $O32$ using calibrations from \citet{Kewley19}. \textit{Panel (d)}: metallicity from $R_{23}$ strong-line diagnostic using calibration from \citet{Curti20a}. \textit{Panel (e)}: metallicity derived from $O3N2$ using calibration from \citet{Marino13}. \textit{Panel (f)}: metallicity derived from the $N2S2H\alpha$ diagnostic as outlined in \citet{Dopita16}. The peak $i$-band flux from SDSS imaging is marked in each panel with a white pentagon. FWHM of the spatial PSF is shown by the red circle in panel (f). The slit shown in panel (f) spans the region of highest S/N for the \OIIIs$\lambda$4363 line and is examined in detail in Fig~\ref{fig:slit_gradient}}
    \label{fig:metallicity_maps}
\end{figure*}

\begin{figure*}
    \centering
    \includegraphics[width=\textwidth]{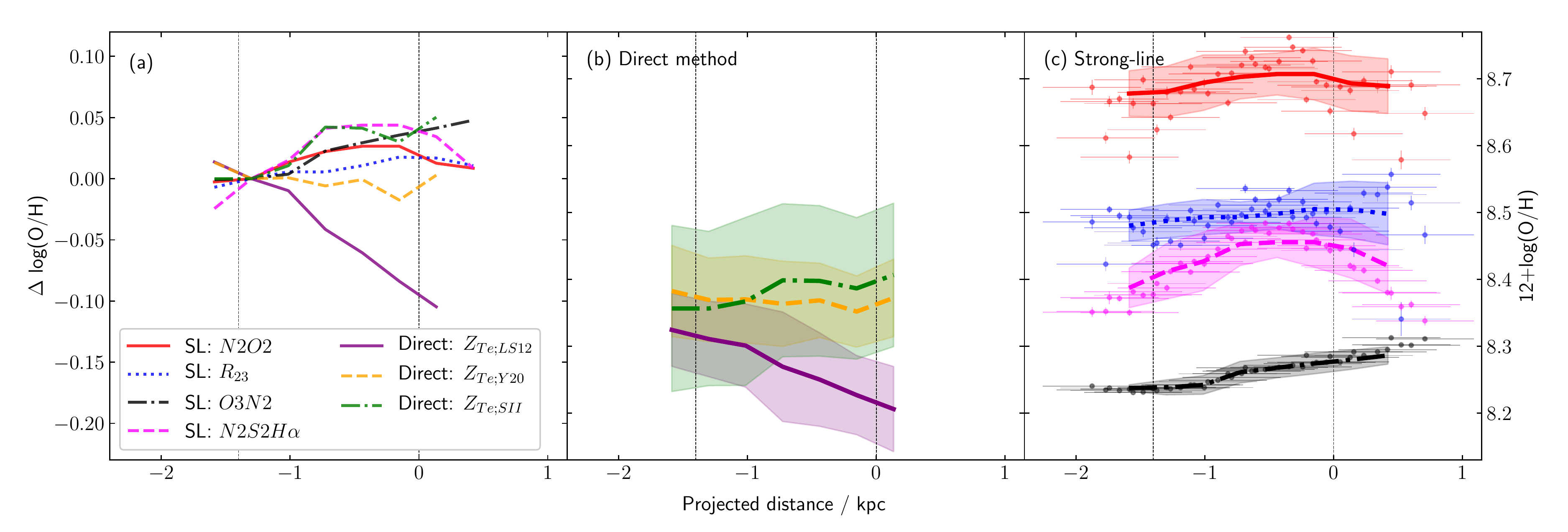}
    \caption{Spatial trend in metallicity along a mock slit for seven different strong-line and direct method metallicity measurement techniques.
    Panel (c) shows individual spaxels and running median trends measured with $N2O2$ (red), $O3N2$ (black), $R_{23}$ (blue) and $N2S2H\alpha$ (magenta). Colour coding is as indicated in the legend in panel (a). More details on these strong-line metallicities can be found in \S~\ref{sub:SL_logOH}.
    Panel (b) reproduces trend lines for three different direct method assumptions from Fig~\ref{fig:direct_metal_map} for ease of comparison.
    Panel (a) renormalises each of these seven trend lines to show metallicity deviation.
    The horizontal axis is zeroed at the adopted core of SAMI609396B, taken as the location of the peak in $i$-band flux from SDSS imaging.
    Vertical error bars show the measurement uncertainty carrying through from emission line measurements. Horizontal error bars indicate the FWHM of the spatial PSF of the SAMI observation in terms of physical distance.
    }
    \label{fig:slit_gradient}
\end{figure*}

\subsection{Strong-Line Metallicity}
\label{sub:SL_logOH}

In Figure~\ref{fig:metallicity_maps} we compare four different strong-line metallicity maps with \ZLS and \ZYates direct method metallicity maps derived in \S \ref{sub:direct_logOH}. Strong-line metallicities are derived using a selection of widely strong-line diagnostics, defined in Equations \ref{eq:line_ratio_N2O2} -- \ref{eq:line_ratio_D16}:

\begin{gather}
    \label{eq:line_ratio_N2O2}
    N2O2 = \text{log}_{10}\left(\text{\NII/\OII}\right)\\
    \label{eq:line_ratio_O32}
    O32 = \text{log}_{10}\left(\text{\OIII/\OII}\right)\\
    \label{eq:line_ratio_R23}
    R_{23} = \text{log}_{10}\left(\frac{\text{\OIIIs}\lambda4959 + \text{\OIIIs}\lambda5007 + \text{\OII}}{H\beta}\right)\\
    \label{eq:line_ratio_N2}
    N2 = \text{log}_{10}\left(\text{\NII/H}\alpha\right)\\
    \label{eq:line_ratio_O3N2}
    O3N2 = \text{log}_{10}\left(\text{\OIII/H}\beta\right) - N2\\
    \label{eq:line_ratio_D16}
    N2S2H\alpha = \text{log}_{10}\left(\text{\NII/\SII}\right) - 0.264 \cdot N2
\end{gather}
\smallskip

where \NIIs = \NIIs$\lambda$6583, \OIIs = (\OIIs$\lambda$3726 + \OIIs$\lambda$3729), \SIIs = (\SIIs$\lambda$6716 + \SIIs$\lambda$6731), and  \OIIIs = \OIIIs$\lambda$5007 unless otherwise specified.
We use strong-line calibrations based on a mixture of theoretical and observational calibrations, outlined as follows:

\begin{itemize}
    \item \textbf{N2O2}: We use the theoretical calibration provided in \citet{Kewley19} to solve iteratively for metallicity and ionisation parameter using the $N2O2$ (Eq.~\ref{eq:line_ratio_N2O2}) and $O32$ (Eq.~\ref{eq:line_ratio_O32}) diagnostic line ratios.

    \item \textbf{R}$_{23}$: We use the calibration provided by \citet{Curti20a} based on direct method measurements of stacked SDSS galaxies. The $R_{23}$ ratio (Eq.~\ref{eq:line_ratio_R23}) is two-valued with a turnover at around 12+log$(O/H)=8.1$. Using $N2$ (Eq.~\ref{eq:line_ratio_N2}) to distinguish between high- and low-metallicity branches, we find $N2>-1.0$ across the extent of SAMI609396B, prompting us to consider only the high-metallicity branch.

    \item \textbf{O3N2}: Calibration based on large compilation of $T_e$ measurements in \HII regions from \citet{Marino13}.

    \item \textbf{N2S2H$\alpha$}: This diagnostic was proposed by \citet{Dopita16} based on predictions from photoionisation modelling. We adopt the calibration presented therein.

\end{itemize}

The colour maps shown in Fig~\ref{fig:metallicity_maps} are shown with different normalisation so as to visualise any spatial trends in metallicity in each diagnostic, setting aside the expected discrepancies in normalisation between alternative diagnostics
\citep[e.g.][]{KewleyEllison08}. Indeed, even after applying the \citetalias{Yates20} correction, the median direct method metallicity ($\widetilde{Z}_{\text{Te; Y20}}=8.40$) is still nearly 0.3 dex lower than that of the theoretically calibrated $N2O2$ diagnostic ($\widetilde{Z}_{\text{N2O2}}=8.68$). This difference is consistent with previous work which has shown systematic offset between metallicities derived from $N2O2$ using theoretical and empirical calibrations \citep{Bresolin09, Bresolin15}.

\subsection{Is The Metallicity Gradient Positive Or Negative?}
\label{sub:gradient}

While it is widely known that different metallicity measurement techniques often disagree in normalisation, one would hope that at a minimum two methods should agree on the ranked order of metallicities they measure.
It is immediately striking from Fig~\ref{fig:metallicity_maps} that even qualitative spatial trends in metallicity are very sensitive to the adopted diagnostic.
Figure~\ref{fig:slit_gradient} illustrates these spatial trends as a 1D projection.
Given the disturbed morphology of SAMI609396B, we do not formally define a metallicity gradient, but instead examine 1D spatial trends along the mock slit shown in Fig~\ref{fig:direct_metal_map} panels (a-c) and Fig~\ref{fig:metallicity_maps} panel (f). This slit encompasses the region of highest emission line signal-to-noise and approximately corresponds to the region of highest $g$-band flux (Fig~\ref{fig:imaging}).

Panel (a) of Fig~\ref{fig:slit_gradient} shows the running medians in metallicity with projected distance along this mock slit for all four strong-line methods described in \S~\ref{sub:SL_logOH} as well as the three different direct method assumptions outlined in \S~\ref{sub:direct_logOH}. The distance axis has been zeroed at the location of peak $i$-band flux from SDSS imaging which we adopt as the core of SAMI609396B. Each trend line has been renormalised relative to the metallicity at $r=-1.3$ kpc. We renormalise at this projected distance rather than the core as the three direct method approaches show best agreement in this spatial region (Fig~\ref{fig:slit_gradient} Panel b). In particular, the \citetalias{Yates20} empirical corrections are smallest in this region.

Most striking in Fig~\ref{fig:slit_gradient} Panel (a) is the clear discrepancy between the \ZLS direct method and all other methods.
The \ZLS method favours a strong trend of decreasing metallicity left-to-right from negative projected distance toward the core. Strong-line methods show an opposite trend, with metallicity increasing in the same direction albeit with less overall deviation from uniform.
As outlined in \S~\ref{sub:direct_logOH}, we find that the \ZYates and \ZSII direct methods both show a much flatter metallicity trend than the \ZLS method, and are in better agreement with strong-line methods.

Given that strong-line methods have their own unsettled systematic uncertainties (Section~\ref{sub:finer_trends}), we do not assess the absolute correctness of `gradients' derived from each method. Instead, we discuss below the physical reason for why the gradient from the \ZLS method is at odds with \ZYates and \ZSII and the strong line methods.

\subsection{O$^{2+}$/O$^{+}$ abundance ratio variation}
\label{sub:o32_variation}

\begin{figure}
    \centering
    \includegraphics[width=\columnwidth]{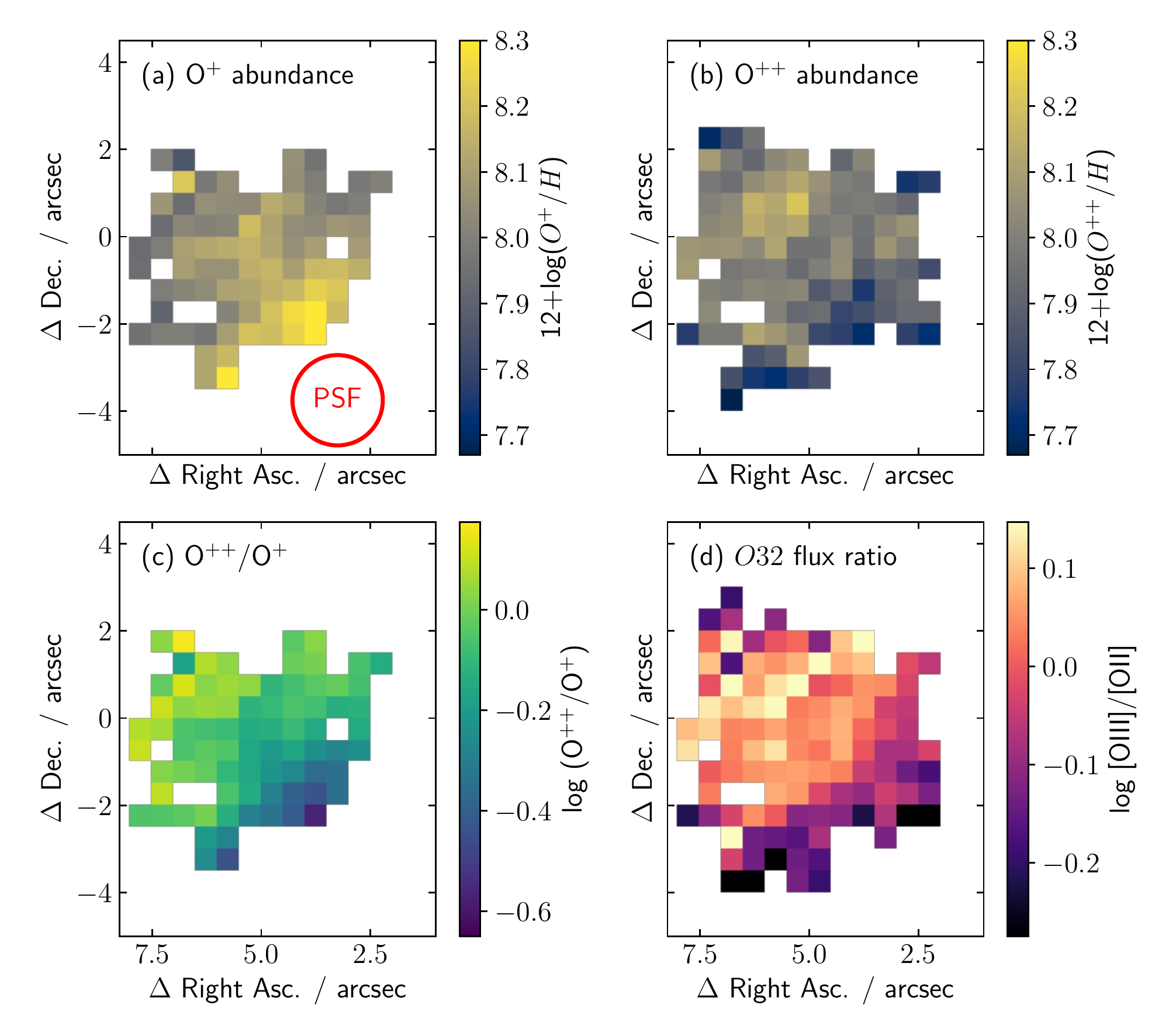}
    \caption{
    Map of derived O$^{++}$/O$^{+}$ abundance ratio for SAMI609396B.
    Panel (a): O$^{+}$/H$^{+}$ abundance derived from Eq~\ref{eq:PM17_O2_abundance} using $T_e$[OII]=$T_e$[SII].
    Panel (b): O$^{++}$/H$^{+}$ abundance derived from Eq~\ref{eq:PM17_O3_abundance} with direct $T_e$[OIII] measurement.
    Note that, unlike in Figures \ref{fig:direct_metal_map} \& \ref{fig:slit_gradient}, abundance maps derived here use maps of $T_e$[SII] and $T_e$[OIII] that have been smoothed by a gaussian filter (FWHM set to measured PSF) to aid in the visual representation of spatial trends.
    Panel (c): O$^{++}$/O$^{+}$ abundance ratio. O$^{+}$ provides a larger contribution to the total oxygen abundance across the majority of SAMI609396B (log(O$^{++}$/O$^{+}) < 0$). Direct metallicities evaluated adopting an assumed $T_e$[OII] -- $T_e$[OIII] relation (e.g. \ZLS in this paper) can underestimate the total oxygen abundance by up to $\sim$0.5 dex in this low ionisation regime (see Fig~7 in \citetalias{Yates20}).
    Panel (d): Observed $O32$ strong-line ratios (Eq~\ref{eq:line_ratio_O32}) appear to correlate with the O$^{++}$/O$^{+}$ abundance ratio when O$^{+}$/H$^{+}$ abundance is derived in this way. Regions of lowest $O32$ correspond to the highest level of correction according to \citetalias{Yates20} correction (see \S~\ref{sub:y20_correction}).
    }
    \label{fig:o32_abundance_ratio}
\end{figure}

We attribute the cause of the discrepancy between \ZLS and other methods to variations in the O$^{2+}$/O$^{+}$ abundance ratio, causing deviations from the fixed $T_e$(\OII) -- $T_e$(\OIII) relation adopted by \ZLSe.
Figure~\ref{fig:o32_abundance_ratio} shows separate O$^+$/H$^+$ and O$^{2+}$/H$^+$ abundance maps, derived using $T_e$(\SII) and $T_e$(\OIII) respectively, with panel (a) showing elevated O$^+$/H$^+$ in the core region (lower-right; corresponding to Projected Distance $\approx0$ kpc in horizontal scale of Fig~\ref{fig:slit_gradient}).

A bulk change in the ionisation structure of \HII regions across SAMI609396B such as this would cause measured temperatures to deviate from the $T_e$(\OII) -- $T_e$(\OIII) relation from \citet{LopezSanchez12} (Eq~\ref{eq:tOII_tOIII}).\footnote{Or, indeed, any fixed monotonic relation assumed between $T_e$(\OII) and $T_e$(\OIII).}
In \S~\ref{sub:te_SII} we noted that $T_e$(\SII) and $T_e$(\OIII) derived for two mock apertures indicated the absence of a strong positive correlation between $T_e$(\SII) and $T_e$(\OIII) (Fig~\ref{fig:tSII_v_tOIII} Panel e). In particular, lower $T_e$(\SII) values obtained in the core region leads to systematically higher O$^{+}$ abundance measurements in \ZSII than \ZLSe, driving the apparent reversal in the measured total oxygen abundance gradient.

Recently, \citet{Yates20} observed that for log(O$^{2+}$/O$^{+}) \lesssim 0.0$, ``semi-direct'' metallicities (that is, metallicities in which $T_e$(\OIII) has been directly measured, but $T_e$(\OII) has been indirectly determined using an assumed $T_e$(\OII) -- $T_e$(\OIII) relation) underestimated the total metallicity by up to $\sim$0.5 dex compared with metallicities derived using direct measurements of both $T_e$(\OII) and $T_e$(\OIII).
This effect also correlates with the \OIII/\OIIs strong-line ratio, motivating the \citetalias{Yates20} correction for observations with log(\OIIIs$\lambda\lambda$4959, 5007 / \OIIs$\lambda\lambda3726, 9) \leq 0.29$.


Figure~\ref{fig:o32_abundance_ratio} shows that O$^{2+}$/O$^{+}$ abundance ratios in SAMI609396B largely fall below log(O$^{2+}$/O$^{+}) \lesssim 0.0$, inside the range highlighted in \citetalias{Yates20} as giving rise to deficits in the total oxygen abundance when ``semi-direct'' methods are used. Furthermore, a spatial trend in O$^{2+}$/O$^{+}$ abundance ratio can be seen in panel (c) of Figure~\ref{fig:o32_abundance_ratio}, with lower O$^{2+}$/O$^{+}$ in the lower-right regions of SAMI609396B. \citetalias{Yates20} found that the ``semi-direct'' abundance deficit is more pronounced at lower values of O$^{2+}$/O$^{+}$. From this, we reason that it is likely that \ZLS underestimates the total oxygen abundance across the majority of SAMI609396B. In particular, the lower O$^{2+}$/O$^{+}$ seen in the core of SAMI609396B indicate that the systematically lower metallicities obtained in the core versus higher radius for \ZLS (panel d of Figure~\ref{fig:direct_metal_map}) can be explained by this semi-direct abundance deficit being amplified in the core region.

By not appropriately accounting for this trend, when applying the \ZLS method the O$^{2+}$/O$^{+}$ abundance ratio trend instead masquerades as the trend in total oxygen abundance seen in Fig~\ref{fig:direct_metal_map} \& \ref{fig:slit_gradient}.

\section{Discussion}
\label{sec:discussion}

\begin{figure*}
    \centering
    \includegraphics[width=\textwidth]{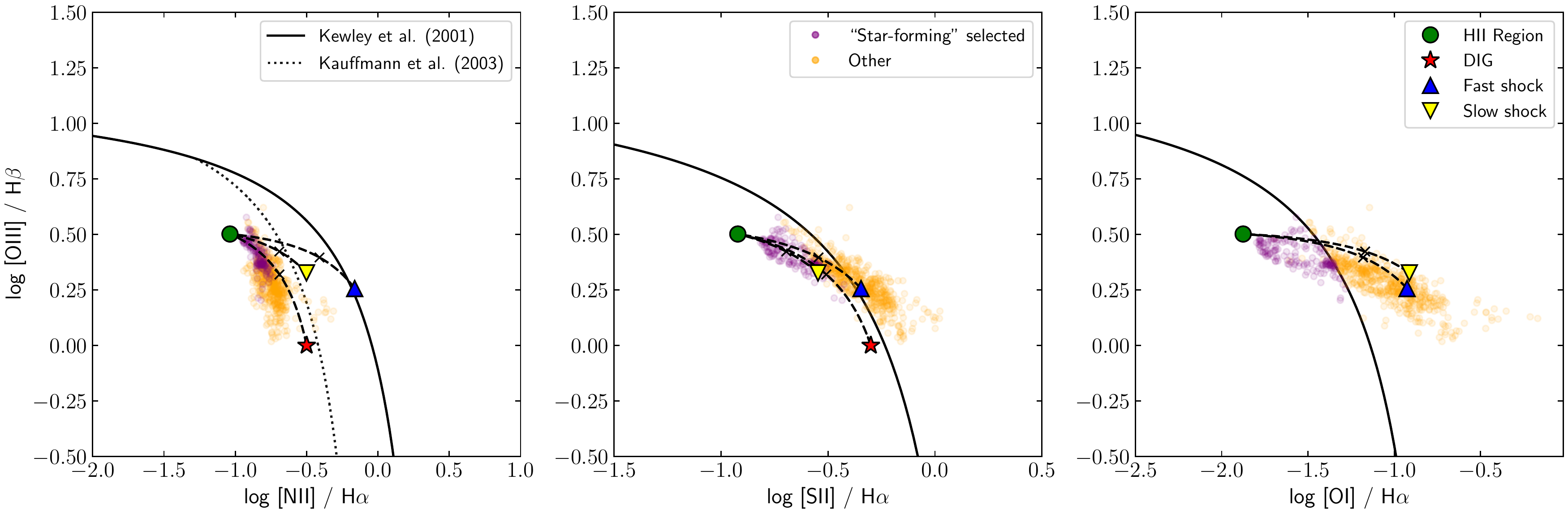}
    \caption{BPT \& VO87 diagnostics diagrams for SAMI609396B. Line ratios for individual spaxels are shown as orange and purple points. \citet{Kewley01} and \citet{Kauffmann03} demarcation lines are shown as solid and dotted grey lines respectively. Purple points denote spaxels below these demarcation lines in each panel. Solid shapes are basis points predicted from photoionisation modelling for \HII regions (\citealt{Dopita13}; green circles), fast shocks (\citealt{Allen08}; blue triangle) and slow shocks (\citealt{Sutherland17}, \citealt{Dopita17}; yellow inverted triangle) according to the model parameters given in Table~\ref{tab:basis_points}. The DIG basis point (red stars) is adopted as the peak region of strong-line ratios from the 10\% lowest surface brightness spaxels in the \citet{Zhang17} MaNGA sample \citep{Sanders17}. Black dashed lines indicate fractional mixing sequences between these basis points.}
    \label{fig:bpt}
\end{figure*}

\subsection{Finer metallicity trends from strong lines}
\label{sub:finer_trends}

The measurement uncertainties on direct method metallicities for SAMI609396B are too large to be used for anything more than the bulk trend.
While the strong-line methods show general agreement when considered in this bulk fashion, deviations exist in the finer details of their spatial trends (Fig~\ref{fig:metallicity_maps} panels c-f \& Fig~\ref{fig:slit_gradient} panel c).
Most notable is the tendency of $O3N2$ to continue to increase beyond the core ($r>0$ kpc in Fig~\ref{fig:slit_gradient}), out to the boundary of the star-formation selected region.
While other strong line methods, especially $N2S2H\alpha$ and $N2O2$, favour a peak in metallicity around $r=-0.6$ kpc and decreasing past the core and beyond.
We explore the possibility of this tension as arising from contaminating emission from non-star-forming sources below in \S~\ref{sub:bpt}.

\subsection{Dissecting the Emission Line Excitation Mechanisms on the BPT Diagram}
\label{sub:bpt}

\begin{table}
    \caption{Input parameters for basis points shown in Figure~\ref{fig:bpt}.}
    \label{tab:basis_points}
    \begin{tabular}{lcccc}
        \hline
        \hline
         & &  $Z/Z_\odot$ & log($q$) & $\kappa$ \\
        \HII region $^a$ & & 1.0 & 7.75 & 50 \\
        \hline
         & $v$ (km s$^{-1}$) & $Z/Z_\odot$ & $n$ (cm$^{-3}$) & $B$ ($\mu$G)\\
        Fast shock $^{b, d}$ & 250 & 1.0 & 10 & 10\\
        Slow shock $^{c, d}$ & 160 & 1.0 & 1000 & 6.1\\
        \hline
        \hline
        \multicolumn{5}{l}{$^a$\citet{Dopita13}; $^b$\citet{Allen08}}\\
        \multicolumn{5}{l}{$^c$\citet{Sutherland17}}\\
        \multicolumn{5}{l}{$^d$Shock basis points include 50\% contribution from pre-cursor}
    \end{tabular}
\end{table}

\begin{figure}
    \centering
    \includegraphics[width=\columnwidth]{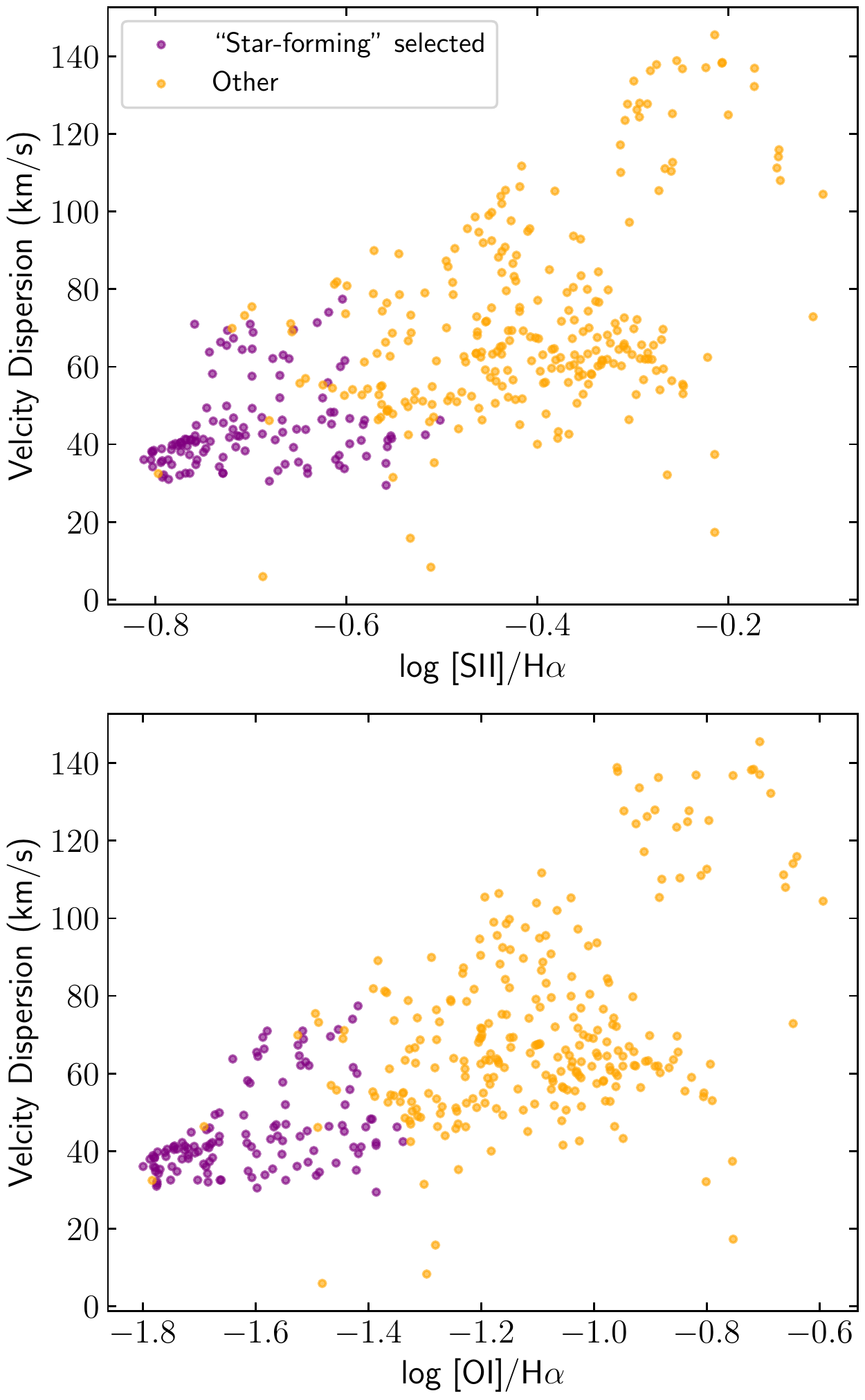}
    \caption{\SII/\Has and \OI/\Has diagnostic line ratios plotted against velocity dispersion for the full SAMI609396 field of view. The positive correlation observed between each of these diagnostic line ratios and velocity dispersion indicates the presence of shocks. The line ratio shown on the horizontal axis is \SIIs$\lambda\lambda$6716, 31 / \Has for the top panel and \OIs$\lambda$6300 / \Has in the bottom panel. Colour coding is as for Figure \ref{fig:bpt}. Emission line fluxes and velocity dispersions shown in this figure are from 1-component fits provided in the SAMI DR2 value-added data products.}
    \label{fig:s2o1_vd}
\end{figure}

Gas-phase metallicity studies such as this aim to determine abundances of nebulae photoionised by recently formed O- and B-type stars (\HII regions). However, emission from other sources including active galactic nuclei (AGN), shock-heated gas (shocks), and diffuse ionised gas (DIG), may contribute significantly to an observed extragalactic emission spectrum.
Since each of these sources exhibit characteristically different emission spectra, inference of the properties of ionised gas from an emission line spectrum requires knowledge (or an assumption) of the excitation mechanism causing the emission.

Different excitation sources are generally distinguished with BPT or VO87 diagnostic diagrams which compare \OIII/H$\beta$ to each of \NII/\Ha, \SII/\Ha, and \OI/\Has \citep{BPT81, VO87}. Demarcation lines that separate \HII regions from other sources of emission have been derived from photoionisation modelling \citep{Kewley01} and from large samples of observational data \citep{Kauffmann03}. These can be used to exclude observations which are dominated by emission sources other than \HII regions.

Of course, the presence of one emission source in an observation does not preclude the presence of any others. Indeed a so-called ``mixing sequence'' is often observed on diagnostic diagrams, spanning the regions between the loci inhabited by \HII regions and those of other ionizing sources.
Global spectra residing along this sequence are best explained as galaxies for which the global spectrum contains emission from both \HII regions and either AGN or shocks, with the position along this mixing sequence determined by the relative proportion of each of these sources of emission.
Further, when observations are made with IFU spectroscopy, mixing sequences can be spatially resolved within individual galaxies \citep{Ho14, Davies14a, Davies14b, Davies16, Davies17, Jones17, Zhang17, DAgostino18} due to differing spatial distributions of emission sources within these galaxies.

Figure~\ref{fig:bpt} shows diagnostic line ratios for individual spaxels from SAMI DR2 single component emission line fits over the full extent of the SAMI609396 merger system. Purple points are those which pass the \HII region  \citet{Kewley01} selection criteria in all three panels. The spatial region selected as SAMI609396B analysed in this paper is a subset of these purple points (refer to Fig~\ref{fig:sami_dr2} for SAMI609396B spatial selection).

Overplotted on Fig~\ref{fig:bpt} are basis points predicted from photoionisation modelling for \HII regions (\citealt{Dopita13}; green circles), fast shocks (\citealt{Allen08}; blue triangle) and slow shocks (\citealt{Sutherland17}, \citealt{Dopita17}; yellow inverted triangle) as well as observed loci of DIG-dominated regions (\citealt{Zhang17, Sanders17}; red star).
The adopted model parameters for each of these basis points are summarised in Table~\ref{tab:basis_points}. Note that the shock model basis points include a contribution from precursor emission. We assume a 50:50 contribution from the shock and precursor. \HII region model parameters are based on metallicity and ionisation parameter values obtained from $N2O2$ and $O32$ line ratios (see \S~\ref{sub:SL_logOH}).
Shock model parameters are difficult to constrain as they are degenerate with fractional contribution and spatial variations, not to mention the large modelling uncertainties. Selected shock velocities (Table~\ref{tab:basis_points}) broadly reflect velocity dispersions observed in SAMI609396 (see Fig~\ref{fig:sami_dr2}~\&~\ref{fig:s2o1_vd}) and were chosen on the basis of how well they visually reproduced the individual points in Figure~\ref{fig:bpt}.
Black dashed lines show mixing models between \HII regions and each of these other emission sources. These lines indicate the sequence that arises by varying in the fractional contribution between the two fixed basis points. The mid-point of each sequence is labelled with a black cross.

In addition to emission line ratios, velocity dispersion is a useful tool for identifying the presence of shocks.
Emission from shocks often shows a positive correlation between velocity dispersion and \SII/\Has or \OI/\Has diagnostic line ratios \citep{Ho14}, while DIG emission will not yield such a correlation.
In Figure~\ref{fig:s2o1_vd}, \SII/\Has and \OI/\Has emission line ratios from SAMI609396 are plotted against measured velocity dispersion, supplementing our BPT and VO87 diagrams. Figure~\ref{fig:s2o1_vd} shows that both \SII/\Has and \OI/\Has ratios are positively correlated with velocity dispersion in SAMI609396. While emission line ratios alone cannot definitively distinguish between emission from shocks and DIG (Fig~\ref{fig:bpt}), on the basis of Figure~\ref{fig:s2o1_vd} we conclude that the dominant source of non-star-forming emission observed in the SAMI609396 data cube is shock-heated gas.

\subsubsection{Effect of contaminating emission}

Given the limited ($\sim$kpc) spatial resolution of SAMI, some amount of contamination from non-star-forming emission sources is inevitable, despite limiting our analysis to the region of nominally star-forming dominated emission.
\citet{Sanders17} showed that contamination from DIG can lead to discrepancies in measured metallicity of up to $\sim$0.3 dex.
In resolved studies, \citet{Poetrodjojo19} found that the inclusion of DIG in metallicity gradient measurements affects all diagnostics to varying degrees.

Of particular concern to establish the robustness of gradient studies is the presence of significant systematic variation in the relative contribution of \HII region and non-star-forming emission. This has the potential to affect the inference on spatial metallicity trends.
Figure~\ref{fig:bpt} suggests that spaxels in this star-forming selected region may form the beginning of a spatial mixing sequence, perhaps indicating existence of spatial variations in the fractional contribution of shock emission to the total emission.
Given the multiple ways metallicities from different diagnostics can be affected by contaminating emission, these variations could help to explain differences in the apparent metallicity trends recovered.

Line ratios plotted in Figures \ref{fig:bpt} \& \ref{fig:s2o1_vd} support our assumption that the ``star-forming'' selected spaxels associated with SAMI609396B are indeed dominated by emission from \HII regions. However, it should be considered that even in regions with emission ``dominated'' by \HII regions, some amount of non-star-forming emission will invariably be present. In particular, the mixing sequences shown as black dashed lines in Figure~\ref{fig:bpt} highlight that there is room for variation in the relative contribution of different emission sources without moving outside the scope of what can be considered ``dominated'' by \HII regions.
A quantitative assessment of this effect is beyond the scope of this paper, but we note that variable contributions of non-star-forming emission in IFU observations of galaxies has the potential to affect measured trends in gas-phase abundances.


In \S~\ref{sec:OH_trends} we showed that, aside from the \ZLS application of the direct method, our metallicity measurements favour a flattened metallicity gradient.
This flat gradient is likely due to the effects of the merger, which are known to produce flattened metallicity gradients due to strong inflows of pristine galaxies from the outskirts of galaxies \citep[e.g.][]{Kewley10}.
The measured gradient may be affected by the presence of shocks, however given that these metallicities were derived using a relatively small subset of the mixing sequence seen in Fig~\ref{fig:bpt} (i.e. the purple points) the effect of this contribution is likely not too significant.




\section{Conclusion} \label{sec:conclusion}

Following a search of the SAMI Galaxy Survey Data Release 2 Public Data, we identified SAMI609396B, an interacting galaxy showing high $S/N$, spatially-resolved detections of three auroral lines: \OIIIs$\lambda$4363, \SIIs$\lambda\lambda$4069, 76 and \SIIIs$\lambda$6312. The source also has properties that make it a good candidate for a local analog of high redshift galaxies, in particular for its combination of moderate stellar mass, disturbed morphology and elevated specific star formation rate (see \S~\ref{sub:value_added} \& Appendix \ref{ap:global_properties}).

We use \OIIIs and \SIIs auroral-to-strong line ratios to derive spatially resolved electron temperature measurements for two sub-regions within the emitting \HII regions ($T_e$(\OIII) and $T_e$(\SII)).
Our results indicate the absence of a strong positive correlation between the $T_e$(\SII) and $T_e$(\OIII) temperatures across different spatial regions in SAMI609396B. Instead, Figure~\ref{fig:tSII_v_tOIII} shows $T_e$(\SII) and $T_e$(\OIII) appearing to trend in opposite directions between two apertures. This deviates from the common assumption of a fixed positive monotonic relation between these different temperatures.

Our $T_e$(\OIII) measurements allow for direct method O$^{2+}$/H$^+$ abundance measurements. We then derive direct method total oxygen abundances under three different treatments of the O$^{+}$/H$^+$ abundance:
\begin{enumerate}
    \item $Z_\text{Te; LS12}$: $T_e$(\OII) is assumed from $T_e$(\OII) -- $T_e$(\OIII) relation \citep{LopezSanchez12}.

    \item $Z_\text{Te; Y20}$: As for $Z_\text{Te; LS12}$, with additional \citetalias{Yates20} empirical correction, based on \OIII/\OII strong-line ratio.

    \item $Z_\text{Te; SII}$: $T_e$(\OII) adopted as $T_e$(\OII) = $T_e$(\SII).
\end{enumerate}

We show that the disagreement between spatial metallicity trends returned by these methods is pronounced. \ZLS favours a strong spatial trend with much lower total oxygen abundances being measured in the core, while \ZYates and \ZSII instead suggest a flatter spatial trend, if anything perhaps opposite to the \ZLS trend.
We conclude that the cause of this disagreement is variation in the O$^{2+}$/O$^{+}$ abundance ratio causing deviations from the assumed $T_e$(\OII) -- $T_e$(\OIII) relation. Accordingly, \ZLS results in systematically lower O$^{+}$ abundances across the whole of SAMI609396B than those of \ZSIIe.
This gives rise to an apparent metallicity gradient as the effect is not spatially uniform: O$^{+}$ abundance is particularly elevated in the core when probed by \ZSIIe.
The measured variation in the O$^{2+}$/O$^{+}$ abundance ratio correlates with variations in the \OIII/\OIIs strong line ratio.
Thus, applying the empirical correction from \citet{Yates20} (\ZYatese) results in a trend more in line with \ZSIIe.
Additionally, we derive metallicity with four strong-line diagnostics ($R_{23}$, $N2O2$, $O3N2$ and $N2S2H\alpha$) using a mixture of observation- and theory-based calibrations. Spatial trends recovered by these strong-line methods again favour opposite trends to that of \ZLS, much more in line with those observed with \ZSII and \ZYates.

From diagnostic diagrams, we identify the presence of non-star-forming emission in the SAMI609396 system. We attribute this emission to shock-heated gas on the basis of the observed correlation between the \SII/\Has emission line ratio and the measured velocity dispersion.
Despite applying our analysis to the star-forming selected region around SAMI609396B, we note that in reality each spaxel will contain some amount of contaminating, non-star-forming emission.
In particular, we show that spaxels in this star-forming selected region appear to form the beginning of a spatial mixing sequence, indicating spatial variations in the fractional contribution of non-star-forming emission to the total emission.
Given the different ways metallicities from different diagnostics can be affected by contaminating emission, these variations could help to explain differences in the apparent metallicity trends recovered.

Aside from the \ZLS application of the direct method, our metallicity measurements favour a flat metallicity gradient for SAMI609396B. This flat gradient can be explained by the effects of the merger which are known to produce flattened metallicity gradients due to inflow of pristine gas from large radii \citep{Kewley10}. However, possible contamination from shock emission may affect the gradient measurement.

The direct method remains the main calibration baseline for studying the chemical evolution of galaxies. However, it is not immune to modelling uncertainties. This study highlights the importance of adequately constraining the internal ionisation and temperature structure within \HII regions when probing spatial variations of the metallicity across galaxies. We have shown here that abundance measurements based on $T_e$(\OIII) alone are not a good indicator of the metallicity gradient in SAMI609396B due to their sensitivity to the ionisation parameter.

Spatially resolved applications of the direct method are currently limited even within the local Universe.
Low-mass galaxies ($<$10$^{9.5}$ $M_\odot$) contribute significantly to the stellar mass density and escape fraction of  hydrogen ionizing photons at high redshift. However, the internal chemical distribution of these low-mass galaxies are rarely constrained owing to the spatial resolution and detection limit.
This situation will be improved by forthcoming facilities such as \emph{JWST}/NIRSpec and ground-based ELTs, which will push both the depth and spatial resolution attainable for IFU observations. In-depth analysis of local objects like SAMI609396B, thus set the stage for future detailed metallicity analysis of low-mass galaxies at high redshift.

\section*{Acknowledgements}

We would like to thank Rob Yates for detailed discussions about the Yates et al. (2020) semi-direct method and for sharing their
insights on using the $O32$ ratio to distinguish between lower and upper branch values in Figure D1.
We are grateful to Tucker Jones for valuable discussions about this work and to Jesse van de Sande and Ned Taylor for sharing SAMI-related expertise. We would also like to thank the referee for their constructive suggestions which certainly improved this work.
This study was based on publicly released data from the SAMI Galaxy Survey.
The SAMI Galaxy Survey is based on observations made at the Anglo-Australian Telescope. The Sydney-AAO Multi-object Integral field spectrograph (SAMI) was developed jointly by the University of Sydney and the Australian Astronomical Observatory.
This research was supported by the Australian Research Council Centre of Excellence for All Sky Astrophysics in 3 Dimensions (ASTRO 3D), through project number CE170100013. AJC acknowledges support from an Australian Government Research Training Program (RTP) Scholarship.

\section*{Data Availability}

This paper uses data from the SAMI Galaxy Survey Public Data Release 2 \citep{Scott18} which are available at \url{https://sami-survey.org/abdr}. Those data products include strong emission line flux maps; the auroral emission line flux maps used here are available from the corresponding author (AJC) on reasonable request.
A list of SAMI galaxies identified in our search as showing auroral line emission is given in Appendix~\ref{ap:auroral_list}.




\bibliographystyle{mnras}
\bibliography{sami_resolved_te} 




\appendix

\section{SAMI Galaxies with Auroral Lines}
\label{ap:auroral_list}

\begin{table}
    \caption{SAMI Galaxies with visually identifiable \OIIIs$\lambda$4363 emission.}
    \label{tab:auroral_list}
    \begin{tabular}{l}
        \hline
         SAMI ID\\
         \hline
         84107\\
         137071\\
         177518\\
         209319\\
         325376\\
         561143\\
         567676\\
         567736\\
         609396\\
        \hline
    \end{tabular}
\end{table}

We provide in Table~\ref{tab:auroral_list} a list of galaxies in SAMI Data Release 2 public data showing visually identifiable \OIIIs$\lambda$4363 emission. We do not claim that this list is exhaustive; rather, it is intended to provide a starting point for any future work hoping to make use of auroral line detections in SAMI data. This list was compiled during an exploratory search of the SAMI Data Release public data by visually inspecting the 1D spectra obtained by binning spaxels with the highest signal-to-noise on the \Has emission line. All cases other than SAMI609396 required some degree of spatial binning to achieve $S/N_{\lambda4363} \gtrsim 5$. In our brief exploration we found that typically fewer than $\sim$4-5 usable bins could be extracted, however we did not expend any effort optimising these binning schemes. Our search focused on the \OIIIs$\lambda$4363 auroral line, however we note that in many of the galaxies listed in Table~\ref{tab:auroral_list}, \SIIs$\lambda\lambda$4069, 76 and \SIIIs$\lambda$6312 are also clearly present. We speculate that there may be galaxies with prominent \SIIs and \SIIIs auroral line emission which were not picked up in our \OIIIs based search.
We note that the \NIIs$\lambda$5755 and \OIIs$\lambda\lambda$7320, 30 auroral lines are typically not observable with SAMI. The \NIIs$\lambda$5755 auroral line falls in the wavelength gap between the blue and red arms in the SAMI datacubes. The \OIIs$\lambda\lambda$7320, 30 line falls near the red limit of the SAMI Galaxy Survey data and often outside the spectral coverage. Even in cases where it falls inside the spectral coverage, we find it is not detectable.

\section{Global properties}
\label{ap:global_properties}

SAMI DR2 value added data products include a spatially resolved star-formation rate (SFR) map based on measured \Has flux (refer to \citealt{Medling18} for details). We derive a global SFR for SAMI609396B by summing the spaxel by spaxel star-formation rate over the SAMI609396B selection mask defined in Section~\ref{sub:value_added}, obtaining SFR $ = 4.21 \pm 0.30 M_\odot$ yr$^{-1}$. However, it is worth noting that the spatial region considered here is limited by the SAMI field-of-view which does not achieve full coverage of SAMI609396B (see Fig~\ref{fig:imaging}). Indeed, the star-formation rate map (Fig~\ref{fig:sami_dr2} panel c) appears to peak near the FoV boundary. It is likely that the region extending beyond the FoV contributes significantly to the global SFR of SAMI609396B. In that sense, we suggest that the quoted SFR can be considered as a lower-bound.
In addition to SAMI609396B, we derive SFR for the more massive companion galaxy by summing the SFR map over the remainder of the SAMI FoV. This yields a value of SFR $ = 0.32 \pm 0.08 M_\odot$ yr$^{-1}$, although we note that this spatial region exhibits significant contribution to its emission spectrum from non-star-forming sources which may bias this value (\S~\ref{sub:bpt}).

We derive global stellar mass values from $g$- and $i$-band photometry using the relation described in Section 4.2 of \citet{Bryant15}, based on stellar mass estimates from \citet{Taylor11}.
We create a deblended segmentation map for the SDSS $g$-band imaging using \texttt{detect\_sources} and \texttt{deblend\_sources} from \texttt{photutils} package \citep{photutils19}. We use default values of 32 multi-thresholding levels and a blending contrast of 0.001 to run the deblending.
The magnitudes obtained from these segmentation images are given in Table~\ref{tab:deblend_mags}.
Applying the \citet{Bryant15} relation to those magnitudes we obtain stellar mass estimates of log($M_*/M_\odot) = 9.11 \pm 0.10$ for SAMI609396B and log($M_*/M_\odot) = 9.78 \pm 0.10$ for its more massive companion. The $0.10$ dex uncertainties reflect the quoted 1-$\sigma$ scatter in this relation \citep{Taylor11}.
We ignore the flux uncertainties from the SDSS imaging as these contribute only 0.001 dex variations in stellar mass.
These values correspond to to a mass ratio of $\sim$0.21 for this merger system
The global SFR and $M_*$ values derived here place SAMI609396B at least 1.3 dex above the star-forming main sequence (SFMS) for local star-forming galaxies \citep{RenziniPeng15}.

\begin{table}
    \caption{Deblended magnitudes used to estimate stellar masses for SAMI609396B and companion}
    \label{tab:deblend_mags}
    \begin{tabular}{lcc}
        \hline
         & $g$-band & $i$-band\\
        \hline
        SAMI609396B & $15.017  \pm 0.001$ & $14.932  \pm 0.001$\\
        Companion & $15.064  \pm 0.001$ & $14.351  \pm 0.001$\\
        \hline
    \end{tabular}
\end{table}

\section{Spectral Fitting}
\label{ap:fitting}

\begin{figure*}
    \centering
    \includegraphics[width=\textwidth]{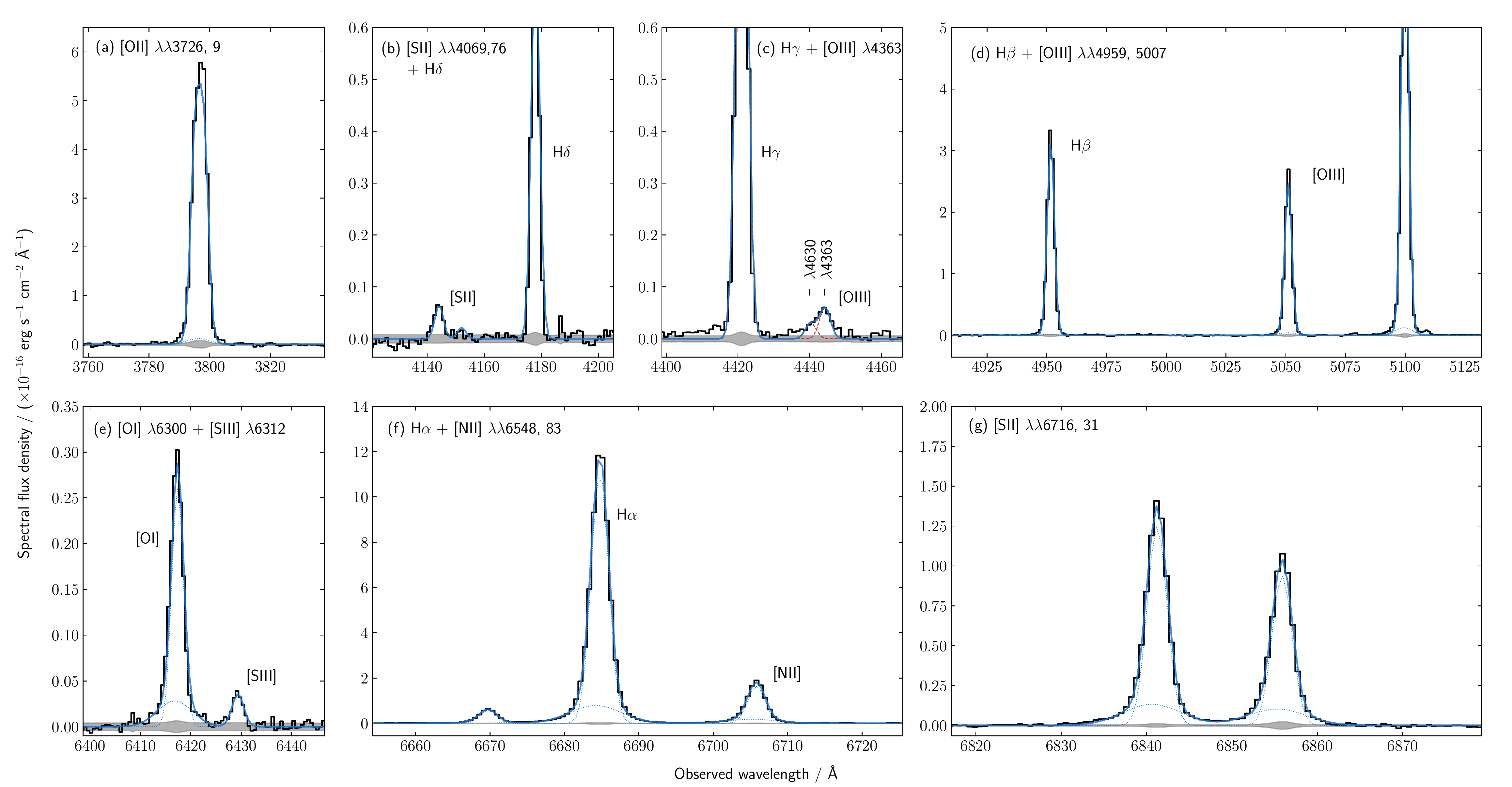}
    \caption{Emission line fits to a 1D spectrum from an individual spaxel. Each panel (a) -- (g) highlights different emission lines used in this analysis as described in the inset text. Units on each axis are the same for each panel, however note the normalisation on the vertical axis varies according to the strength of the emission lines shown. In each panel the continuum subtracted spectrum is shown as the black step plot, while the grey shaded band shows the 1-$\sigma$ error spectrum. The best-fit emission line model is shown by the solid blue line, while the blue dotted lines show the individual broad- and narrow-components fit to the profile, where applicable. The three auroral lines observed in SAMI609396B, [SII]$\lambda\lambda$4069, 76, [OIII]$\lambda$4363, and [SIII]$\lambda$6312, are shown in panels (b), (c), and (e) respectively. The red dashed lines in panel (c) show the individual profiles of the [FeII]$\lambda$4360 and [OIII]$\lambda$4363 emission lines, fit simultaneously to account for blending of these features (see \S \ref{apsub:FeOIIIblending}).}
    \label{fig:spaxel1D}
\end{figure*}

The SAMI DR2 value-added release includes emission line maps for the most widely used strong optical lines. However, the primary focus of this work is the auroral emission lines, which are not included in these data products. Here we outline the methods behind our own spectral fitting to the SAMI DR2 data cubes. Flux measurements obtained here are used throughout \S~\ref{sec:te}-\ref{sec:OH_trends}.

\subsection{Continuum Subtraction}
\label{apsub:continuum}

For each spaxel in the reduced SAMI datacube, we fit the stellar continuum of the blue and red arms simultaneously using \texttt{pPXF} \citep{pPXF17}.
The 1D spectra for each spaxel are logarithmically rebinned. Given the different wavelength resolutions of the two arms, the higher resolution red arm is sampled down to match the velocity scale of the lower spectral resolution blue arm.
Four moment fits are performed to the stellar continuum using the MILES library of stellar templates \citep{Vazdekis10}.
Two moment fits to the Balmer emission lines and strongest forbidden emission lines (\OIII$\lambda\lambda$4959,5007, \NII$\lambda\lambda$6548,83 and \SII$\lambda\lambda$6716,31) are included in the fitting procedure, however these derived emission line fluxes are discarded (refer to \S \ref{apsub:emline_fitting} for emission line fitting).

SAMI609396 exhibits strong emission lines with high equivalent-widths ($>$ 200 \AA{}) and thus even spectra from individual spaxels feature many faint emission lines that are not widely studied.
To ensure that the effect of these faint emission lines on the stellar template fitting is minimised, an iterative sigma-clipping method is employed (described in \S 2.1 of \citealt{Cappellari02}). In this approach, once a global minimum is found, spectral pixels deviating more than 3$\sigma$ from this best-fit template are masked out and a new global fit is obtained. This  process is repeated until no additional pixels are masked out. As a final step, the fits were visually inspected to ensure no spurious spectral features affected the fitting.

A continuum subtracted spectrum was obtained for each spaxel by subtracting the best-fitting stellar spectrum from the reduced observed spectrum from the SAMI data cube.

\subsection{Emission Line Fitting}
\label{apsub:emline_fitting}

Emission line maps of the most commonly used strong emission lines are provided in the value-added SAMI public release data. However, to ensure our emission line ratios are making self-consistent comparisons between strong-lines and the faint auroral lines, we perform new fits to these strong lines as well as the auroral lines. The emission line fitting procedure applied to each spaxel is as follows.

We first fit the \Ha\ emission with a two-component Gaussian profile. We perform a $\chi^2$-minimisation fit across the 6672 -- 6698 \AA{} wavelength range (rest-frame 6553 -- 6578 \AA{} at SAMI609396B catalog redshift). This range encompasses >99\% of the \Ha\ emission for the redshift range covered by the spaxels of this object while minimising contribution from nearby \NII\ emission lines. The \Ha\ emission is well modelled as a primary narrow component (median FWHM$_{\textit{H}\alpha\textit{, nar}} = 2.26$ \AA{}, median redshift $z_{\textit{nar}}=0.018566$) and a secondary broad component (median FWHM$_{\textit{H}\alpha\textit{, brd}} = 7.07$ \AA{}, median redshift $z_{\textit{brd}}=0.018493$) across the spatial extent of SAMI609396B.

We then fix the velocity and velocity dispersion for each of these two kinematic components to values obtained from \Ha\ and simultaneously fit across the full optical wavelength range for broad- and narrow-component fluxes for each of the strong emission lines (\OIIs$\lambda\lambda$3726,9, H$\beta$, \OIIIs$\lambda\lambda$4959,5007, \OIs$\lambda$6300, \NIIs$\lambda\lambda$6548,83, \Ha, \SIIs$\lambda$6716, and \SIIs$\lambda$6731).

The flux of each of these components is allowed to vary freely above a lower-bound of $f_\text{comp} \geq 0$ with the exception of \OIII$\lambda$4959 and \NII$\lambda$6548. The fluxes of each component of these lines are tied to the flux of the corresponding component of \OIIIs$\lambda$5007 and \NIIs$\lambda$6583, respectively, according to following theoretical ratios:
$f_{5007} = 3 \cdot f_{4959}$ and $f_{6583} = 2.9 \cdot f_{6548}$.

We calculate the uncertainty in the flux for each component by adding in quadrature the statistical error from the fit to an estimate of the uncertainty in the level of the continuum. This continuum uncertainty term is calculated as $\sigma_l = \sigma_c \cdot \sqrt{N + EW/\Delta}$ (Eq. 1 in \citealt{PerezMontero17}) where $\sigma_c$ is the standard deviation in a 30 \AA{} range near the emission line, selected to contain only continuum flux, $N$ is the number of spectral pixels encompassed by the fit Gaussian, $EW$ is the equivalent width of the line, and $\Delta$ is the spectral dispersion (\AA{}/pixel).
When considering the `total' emission (i.e. sum of both components), the adopted uncertainty is the uncertainty of both components summed in quadrature.

With the exception of \OIs$\lambda$6300, the faintest of these ``strong-lines'', the fits to these strong lines achieve summed component $S/N > 20$ across the entire spatial extent of interest and achieve $S/N > 50$ in over 95\% of the spaxels. The \OIs$\lambda$6300 line fits achieve $S/N > 20$ in $>$95\% of spaxels but only $S/N > 50$ in the brightest 8 spaxels. These two-component fits were visually inspected, adding confidence to the automatised algorithms.

The fainter emission lines do not present with sufficient signal-to-noise to be reliably modelled with two-component fits. Instead, to these fainter lines we make single component fits where the velocity and velocity dispersion are fixed to those derived for the dominant narrow component above. The faint lines for which single components are used include the three auroral emission lines visually identified to be present (\SIIs$\lambda\lambda$4069,76, \OIIIs$\lambda$4363, and \SIIIs$\lambda$6312), and the fainter Balmer emission lines H$\delta$ and H$\gamma$. Additionally, we include the \FeII\,$\lambda$4360 emission line in our single component fit. 
The effect of the presence of this blended emission feature on the measured \OIII\ flux is discussed in detail in \S \ref{apsub:FeOIIIblending}.

As before, these emission line fluxes are allowed to vary freely with constraint of $f \geq 0$ except that fluxes of the \SIIs$\lambda\lambda$4069, 76 doublet are constrained such that $f_{4069} = 3 \cdot f_{4076}$. Uncertainties for each line are calculated in the same way as described for the individual components of the strong line fits described above. Once derived for each spaxel, these emission line fluxes are collated into 2D maps. Emission line fits for an example 1D spaxel spectrum are shown in Figure~\ref{fig:spaxel1D}.

As a final step, we correct these 2D emission line flux maps using values from the extinction correction map provided in the SAMI DR2 data, derived using spatially smoothed Balmer decrements ($f_{\textit{H}\alpha}/f_{\textit{H}\beta}$) to account for aliasing effects introduced by the SAMI observing process (refer to \citealt{Green18} and \citealt{Medling18} for details), assuming a \citet{Cardelli89} extinction law with $R_V = 3.1$. Unless otherwise specified, the analysis  of this paper is conducted using these reddening corrected emission line flux maps.

\subsection{[FeII]$\lambda$4360 \& [OIII]$\lambda$4363 blending}
\label{apsub:FeOIIIblending}

\begin{figure}
    \centering
    \includegraphics[width=\columnwidth]{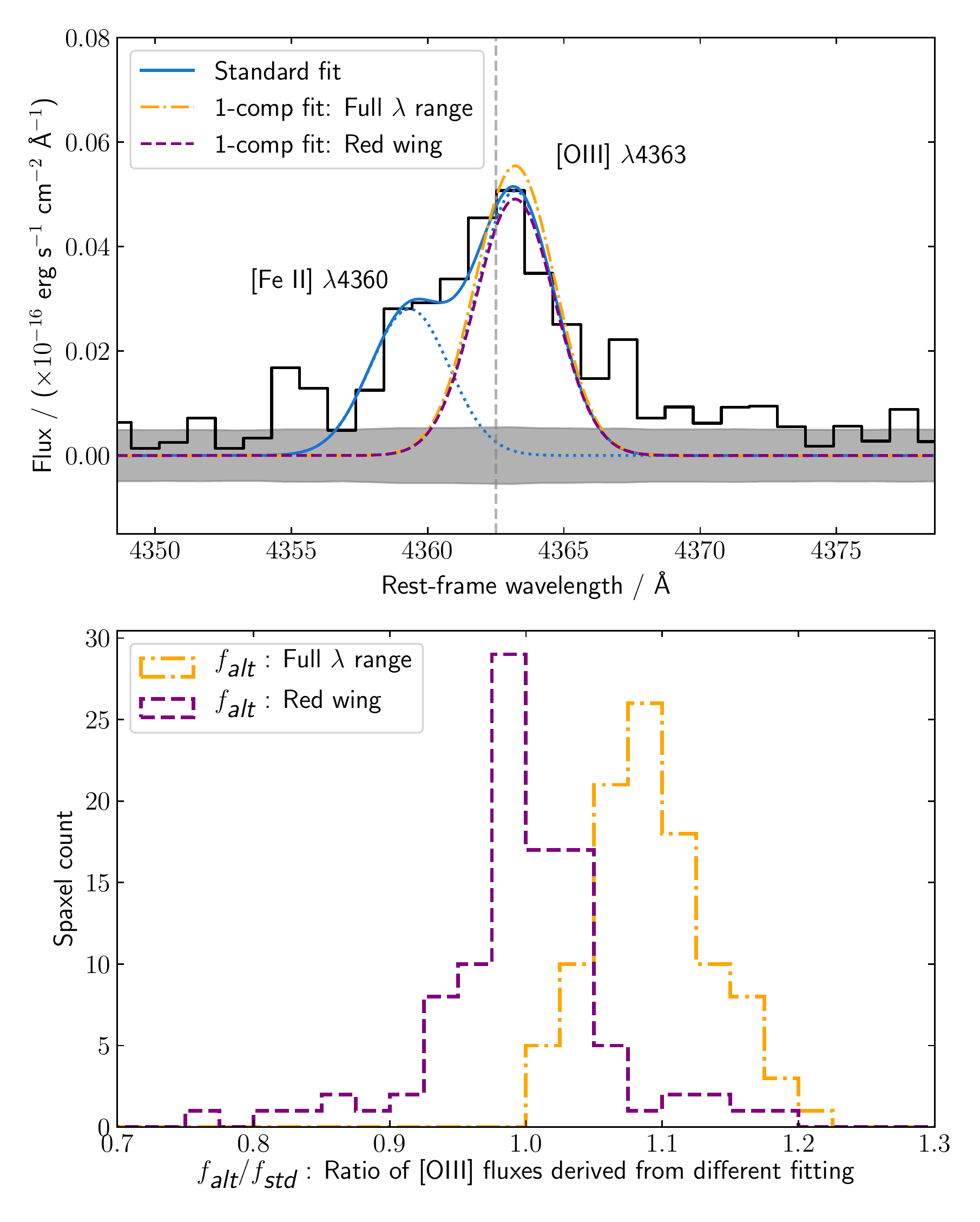}
    \caption{\textit{Top panel:} example of fits obtained to the region around rest-frame $\lambda$4363 when three different fitting methods are implemented. These methods are described in Appendix~\ref{apsub:FeOIIIblending}. Method (i) is shown by blue solid line, with dotted lines showing each component. Method (ii) is shown by the orange dash-dotted line, while method (iii) is shown as the purple dashed line. The wavelength range used for method (iii) is that red-ward of 4362.5 \AA{} (denoted by vertical dashed line).
    \textit{Bottom panel:} Histogram of spaxels with $S/N_{\lambda4363}>3$ of [OIII] flux measurements obtained from methods (ii) and (iii), shown relative to [OIII] flux obtained from method (i). As expected, when [FeII] emission is not accounted for as in (ii), [OIII] flux is systematically higher. While methods (iii) and (i) do not universally agree, the lack of systematic offset gives us confidence that either of these methods on average reliably account for [FeII] emission.}
    \label{fig:FeO_blending}
\end{figure}

Several recent studies have identified an emission feature at $\lambda$4360 \AA{} which may be blended with the \OIII$\lambda$4363 emission line, attributed to an \FeII emission line \citep{Curti17, Berg20, ArellanoCordova20}. From visual inspection of 1D spectra, we identify that \OIII$\lambda$4363 emission often presents with an extended blue wing, which we attribute to blending with this \FeII$\lambda$4360 line.

We account for this by including an emission feature at this wavelength in our line-fitting routine whose flux is allowed to vary freely (Appendix \ref{apsub:emline_fitting}). As with all faint lines in the line-fitting, the velocity and velocity dispersion is tied to that of the narrow component identified for H$\alpha$. Visually, the fits obtained appear to model the emission features around $\lambda$4363 \AA{} well. The median ratio between the \OIII\ and \FeII\ lines across the spatial extent with $S/N_{\lambda4363} > 3$ is $f_{4363}/f_{4360} = 2.1$ with standard deviation $\sigma_{4363/4360} = 1.47$.

As a check of how reliable our \OIII\ flux measurements with this blended \FeII\ + \OIII\ profile are, we fit this wavelength region using three approaches and compare the results. The three approaches are:
\begin{enumerate}
    \item \textit{Standard fitting}: As described in Appendix \ref{apsub:emline_fitting} where \FeII\ and \OIII\ components are simultaneously fit for.

    \item \textit{Naive single component}: A simple single component fit to \OIII\  across the wavelength range from 4345 -- 4380 \AA{}. No attempt is made to account for \FeII\ emission.

    \item \textit{Red wing single component}: A single component is fit to \OIII, excluding pixels blueward of $\lambda = 4362.5 \times (1 + z_\text{fit})$ \AA{},  where $z_\text{fit}$ is the redshift value obtained from the narrow-component fit to \Ha\ for the spaxel in question. This should mask out spectral pixels with $>$ 5 \% contribution from \FeII emission.
\end{enumerate}

In each approach, the velocity and velocity dispersion values are fixed to those obtained for the \Ha\ narrow component, as in our standard fitting.
The best-fit profiles from each of these approaches are shown in the top panel of Figure~\ref{fig:FeO_blending}.
The bottom panel shows the distribution of values obtained for $f_\textit{alt}/f_\textit{std}$ ratios, where $f_\textit{std}$ is the \OIII\ flux obtained from method (i) and $f_\textit{alt}$ is the flux from approaches (ii) and (iii). As expected, when no attempt is made to account for \FeII emission as in method (ii), the \OIII\ flux is systematically overestimated by around 10 \% ($f_\textit{alt}/f_\textit{std} = 1.09 \pm 0.04$; orange dash-dotted line in bottom panel of Figure~\ref{fig:FeO_blending}). However, when applying method (iii), where the blue wing of \OIII\ is masked out, we see no significant systematic offset from our values obtained by method (i) ($f_\textit{alt}/f_\textit{std} = 0.99 \pm 0.07$; purple dashed line in bottom panel of Figure~\ref{fig:FeO_blending}).
We note that there is large scatter in the distribution of $f_\textit{alt}/f_\textit{std}$ for this latter case. This uncertainty of $>$10\% is not unreasonable for observations with $S/N \sim 3-15$. We note also that when using these auroral line fluxes to derive electron temperature measurements, measurement uncertainties of this level are likely outweighed by modelling uncertainties (refer to \S~\ref{sec:te})

From this we conclude that for the \OIIIs/\FeII flux ratios we observe ($f_{4363}/f_{4360} \sim 2.1$), on average methods (i) and (iii) each suffer minimally from contamination by this blended emission feature at $\lambda$4360. \OIII$\lambda4363$ flux measurements quoted in other sections are those derived from the standard fitting routine (method (i)).

\section{Implementing the method of Yates et al. (2020)}
\label{ap:yates_method}

Measurements of electron temperature ($T_e$) are highly sought after in chemical abundance studies as they enable ``direct-method'' metallicity measurements to be made. Full application of the direct method to determine the total oxygen abundance requires measurements of both $T_e$(\OIII) and $T_e$(\OII), such that both O$^{2+}$/H$^+$ and O$^{+}$/H$^+$ can be determined.
However, given the faintness of the required auroral emission lines, often only $T_e$(\OIII) can be measured directly.
In this case, it is common to apply a ``semi-direct'' method.
In this approach, the $T_e$(\OII) is indirectly determined from the measured $T_e$(\OIII) via some assumed relation, allowing the $O^+$ abundance to be derived.

While it is common to assume a simple positive correlation between $T_e$(\OII) and $T_e$(\OIII) \citep[e.g.][]{Izotov06, LopezSanchez12}, \citet{Yates20} (\citetalias{Yates20} hereafter) find that this does a poor job of describing the observed scatter about this relation. Instead, \citetalias{Yates20} highlight that at fixed metallicity, $T_e$(\OII) is anti-correlated with $T_e$(\OIII), and that the general positive trend between $T_e$(\OII) and $T_e$(\OIII) is due to the fact that both correlate negatively with metallicity and in general will both be higher in lower metallicity systems.

Based on these observations, \citet{Yates20} have outlined a new method for determining $T_e$(\OII) and metallicity in systems where only $T_e$(\OIII) can be directly measured.
Unlike previous semi-direct methods, in the \citetalias{Yates20} method, $T_e$(\OII) and metallicity are solved for simultaneously.
This differs from previous semi-direct methods in which $T_e$(\OII) is usually determined based on $T_e$(\OIII) independently of metallicity. Metallicity is then subsequently determined using the value obtained for $T_e$(\OII).

We note that this method (``\citetalias{Yates20} method'') is separate to the empirical correction described in that same publication which we performed in \S~\ref{sub:y20_correction} (``\citetalias{Yates20} correction''), and the \citetalias{Yates20} empirical correction still needs to be applied to the metallicities arising from the \citetalias{Yates20} method.

\begin{figure}
    \centering
    \includegraphics[width=\columnwidth]{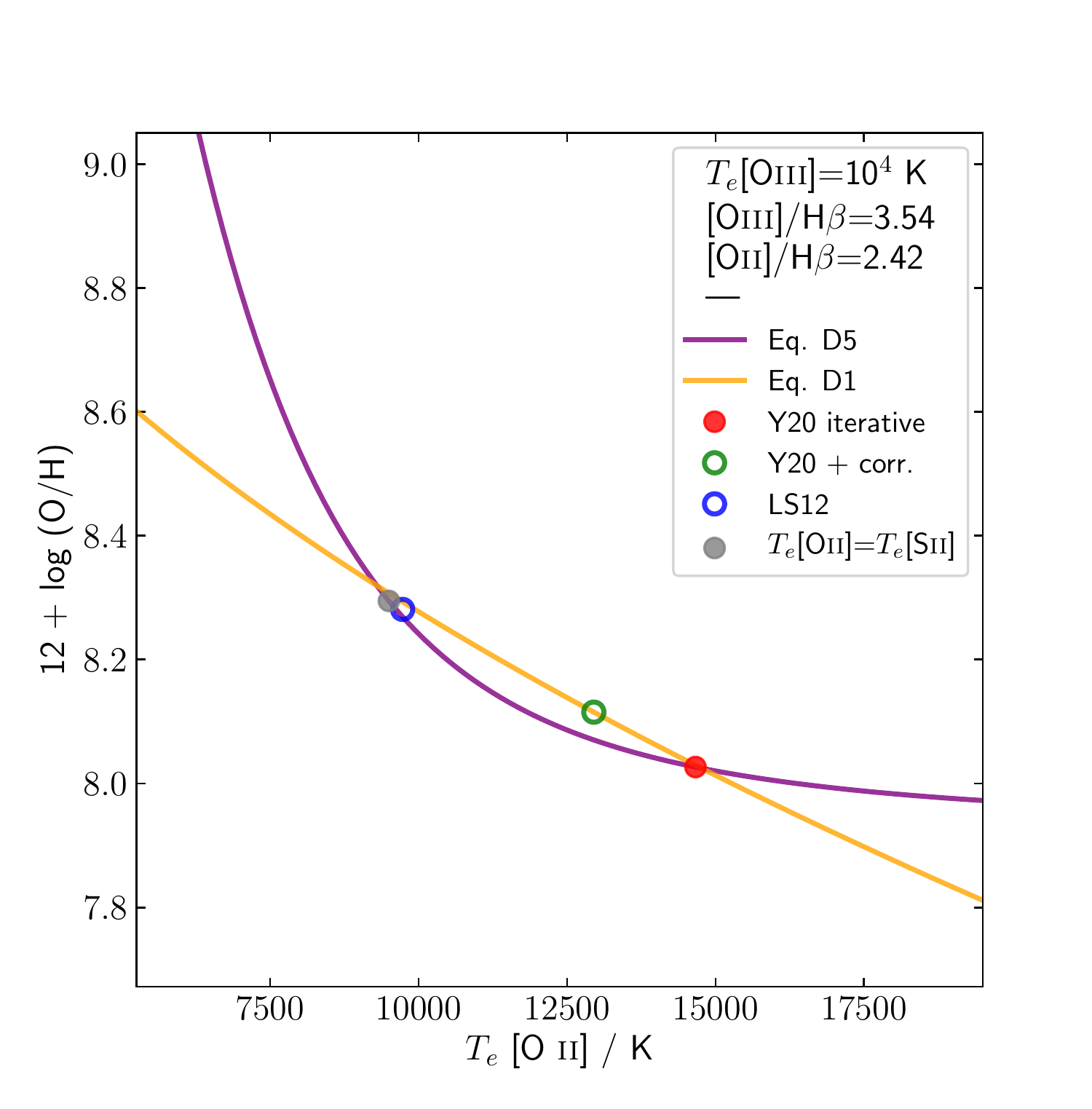}
    \caption{Solving for the intersection of equations \ref{eq:ap_t2t3} and \ref{eq:ap_Zte} from the \citetalias{Yates20} method yields two solutions: an `upper-branch' solution on the left-hand side with a high metallicity, and a `lower-branch' value on the right-hand side with a low metallicity.   Here, the relationship between $T_e$(\OII) and $Z_\text{Te}$ is shown for fixed values of $T_e$(\OIII), \OIII/H$\beta$, and \OII/H$\beta$ measured from a typical spaxel in SAMI609396B (see legend). The purple line shows oxygen abundance determined according to equations \ref{eq:ap_O2} -- \ref{eq:ap_Zte} for various values of $T_e$(\OII), while the orange line shows \citet{Yates20} metallicity-dependent fit to the $T_e$(\OII) -- $T_e$(\OIII) relation at a fixed value of $T_e$(\OIII).
    The \citetalias{Yates20} iterative method as originally outlined favours the lower branch solution (red circle). The \citetalias{Yates20} correction increases the metallicity (green open circle), although not to the value of the upper branch. Both \ZLS and \ZSII (refer to \S~\ref{sec:OH_trends}; $T_e$(\SII$)=9500$ K) fall quite close to the upper branch solution (open blue circle and filled grey circle respectively).}
    \label{fig:ap_y20_two_soln}
\end{figure}

In Section~\ref{sec:OH_trends} we applied a simple semi-direct method to our SAMI609396B data in which $T_e$(\OII) was determined from $T_e$(\OIII) using the calibration of \citet{LopezSanchez12} (Eq.~\ref{eq:tOII_tOIII} in \S~\ref{sub:direct_logOH}) and then the metallicity subsequently determined accordingly (that method was referred to as \ZLS throughout this work). Here, we additionally apply the \citetalias{Yates20} method to our SAMI609396B data. We find the results of the \citetalias{Yates20} method to be double-valued (Figure~\ref{fig:ap_y20_two_soln}). The ``upper branch'' largely agrees with our \ZLS metallicities within $\sim$0.15 dex (median absolute offset is 0.06 dex), however the ``lower branch'' gives starkly different values.
Spatial metallicity trends arising from each branch of this method differ noticeably (Figure~\ref{fig:ap_y20_LS12_compare}),  making it difficult to draw conclusions without further characterisation of the behaviour of each branch in a larger data set.
We describe the details of our implementation of the \citetalias{Yates20} method below.

\subsection{Basis of the Yates Method}
\label{ap:y20_1intro}

The \citetalias{Yates20} method differs from other semi-direct methods in that $T_e$(\OII) and metallicity ($Z_\text{Te}$) are evaluated simultaneously in order to account for the interdependence of these parameters at fixed $T_e$(\OIII); whereas typically semi-direct methods have involved inferring a $T_e$(\OII) value from a $T_e$(\OIII) measurement via a fixed relation and then subsequently determining the metallicity using this value.

The \citetalias{Yates20} method centres on a metallicity-dependent fit to the $T_e$(\OII) -- $T_e$(\OIII) relation, outlined as follows:

\begin{equation}
\label{eq:ap_t2t3}
    T_e(\text{\OII}) = \frac{a(Z_\text{Te})^2}{2 \cdot T_e(\text{\OIII})},
\end{equation}{}

where

\begin{equation}
\label{eq:ap_aZTe}
    a = -12030.22 \cdot Z_\text{Te} + 113720.75.
\end{equation}{}

This can be solved simultaneously with the following equations which determine oxygen abundance from measured \OII/H$\beta$ and \OIII/H$\beta$ line ratios given values for $T_e$(\OII) and $T_e$(\OIII):

\begin{multline}
    \label{eq:ap_O2}
    O^+/H^+ = \frac{\text{\OII}\lambda\lambda3726, 29}{\text{H}\beta} ~ g_1 ~ \alpha_{\text{H}\beta} ~ \sqrt{T_e\text{\OII}}\\ \times \text{exp}[E_{12}/kT_e(\text{\OII})] \times \frac{\beta}{E_{12}\Upsilon_{12}}
\end{multline}

\begin{multline}
    \label{eq:ap_O3}
    O^{++}/H^+ = \frac{\text{\OIII}\lambda\lambda4959, 5007}{\text{H}\beta} ~ g_1 ~ \alpha_{\text{H}\beta} ~ \sqrt{T_e\text{\OIII}}\\ \times \text{exp}[E_{12}/kT_e(\text{\OIII})] \times \frac{\beta}{E_{12}\Upsilon_{12}}
\end{multline}

\begin{equation}
    \label{eq:ap_Zte}
    Z_\text{Te} \equiv 12 + \text{log}(O^+/H^+ + O^{++}/H^+).
\end{equation}
\smallskip

The reader is referred to \citet{Yates20} and \citet{Nicholls14} for more details on these equations including the values and calculations of  various parameters used.

\subsection{Two-valued solution of the Yates method}
\label{ap:y20_2twoval}

In \citet{Yates20}, the authors propose solving these equations with fixed point iteration, however we instead propose numerically solving for the intersection of equations~\ref{eq:ap_t2t3}~\&~\ref{eq:ap_Zte}. This preference is based on our observation that the relations described in Equations~\ref{eq:ap_t2t3} \& \ref{eq:ap_Zte} in fact yield two solutions within the range of what could be considered physically reasonable.\footnote{More generally, it is possible that for some observations there will be no solution. In these cases, solving via the iterative method may be preferable (Yates 2020; private communication)}

This is illustrated in Figure~\ref{fig:ap_y20_two_soln} for an example typical SAMI609396B spaxel with $T_e$(\OIII) = $10^4$ K, and \OIII/H$\beta$ and \OII/H$\beta$ line ratios of 3.54 and 2.42 respectively. For this example spaxel it can be seen that two possible solutions exist: (1) at $T_e$(\OII) = 9,342 K and $Z_\text{Te} = 8.31$, and (2) at $T_e$(\OII) = 14,662 K and $Z_\text{Te} = 8.03$. Neither of these solutions is physically implausible and while solution (1) would fall in a more densely populated region of the $T_e$(\OII) -- $T_e$(\OIII) relation as shown in \citet{Yates20} (refer to Figure 5 in that paper), observed points comparable to solution (2) are found in their sample too.

Indeed, beyond the single example shown in Figure~\ref{fig:ap_y20_two_soln}, we find that all spaxels in SAMI609396B with at least one solution have precisely two.
How then should we decide which of these two solutions to adopt?

The blue open circle in Figure~\ref{fig:ap_y20_two_soln} shows that the $T_e$(\OII) obtained via the \citet{LopezSanchez12} $T_e$(\OII) -- $T_e$(\OIII) relation (refer to Eq. \ref{eq:tOII_tOIII} in \S~\ref{sub:direct_logOH}) agrees quite well with the lower valued $T_e$(\OII) solution. However, this disregards the point of the \citetalias{Yates20} method and findings presented in this work: that assuming a simple fixed relationship between $T_e$(\OII) and $T_e$(\OIII) can be misleading.

The \citetalias{Yates20} iterative method as originally applied in \citet{Yates20} selects for the ``lower branch'' (red circle in Fig~\ref{fig:ap_y20_two_soln}). However, subsequent application of the \citetalias{Yates20} correction serves to shift this point to a slightly higher metallicity, partly toward the upper branch solution (green open circle in Fig~\ref{fig:ap_y20_two_soln}).

To investigate this further, the original \citetalias{Yates20} sample was revisited with the two-valued solution in mind (Yates 2020; private communication).
It was found that targets where the direct\footnote{Here ``direct'' metallicity refers to a metallicity in which both $T_e$(\OII) and $T_e$(\OIII) have been directly measured with auroral lines.} metallicity was closer to the ``upper branch'' solution were often targets with an $O32$ value below the threshold value for which the \citetalias{Yates20} correction should be applied ($O32 \leq 0.29$).
Thus, this $O32$ threshold could be used to distinguish between the lower and upper branches. All spaxels across the spatial extent of SAMI609396B fall in this category with $O32 \leq 0.29$ (refer to Fig~\ref{fig:o32_abundance_ratio}), meaning we would adopt the upper branch value under this scheme, rather than the lower branch value favoured by the original iterative implementation.

While we do not directly measure $T_e$(\OII), in \S~\ref{sub:te_SII} we derived $T_e$(\SII) from the \SIIs $\lambda\lambda$4069, 76 / $\lambda\lambda$6716, 31 line ratio (refer to Figure \ref{fig:tSII_v_tOIII}). Panel (e) of Figure~\ref{fig:tSII_v_tOIII} shows that $T_e$(\SII) < $T_e$(\OIII) across the majority of spaxels in SAMI609396B, with a median value of $T_e$(\SII) = 9,295 K. Given that previous studies have found that $T_e$(\SII) and $T_e$(\OII) are often in general agreement \citep[e.g.][]{Croxall16}, we consider that this additionally supports our selection of the upper-branch value. We note, however, that large scatter is known to exist in both the $T_e$(\SII) -- $T_e$(\OII) and $T_e$(\SII) -- $T_e$(\OIII) relations.

\begin{figure*}
    \centering
    \includegraphics[width=\textwidth]{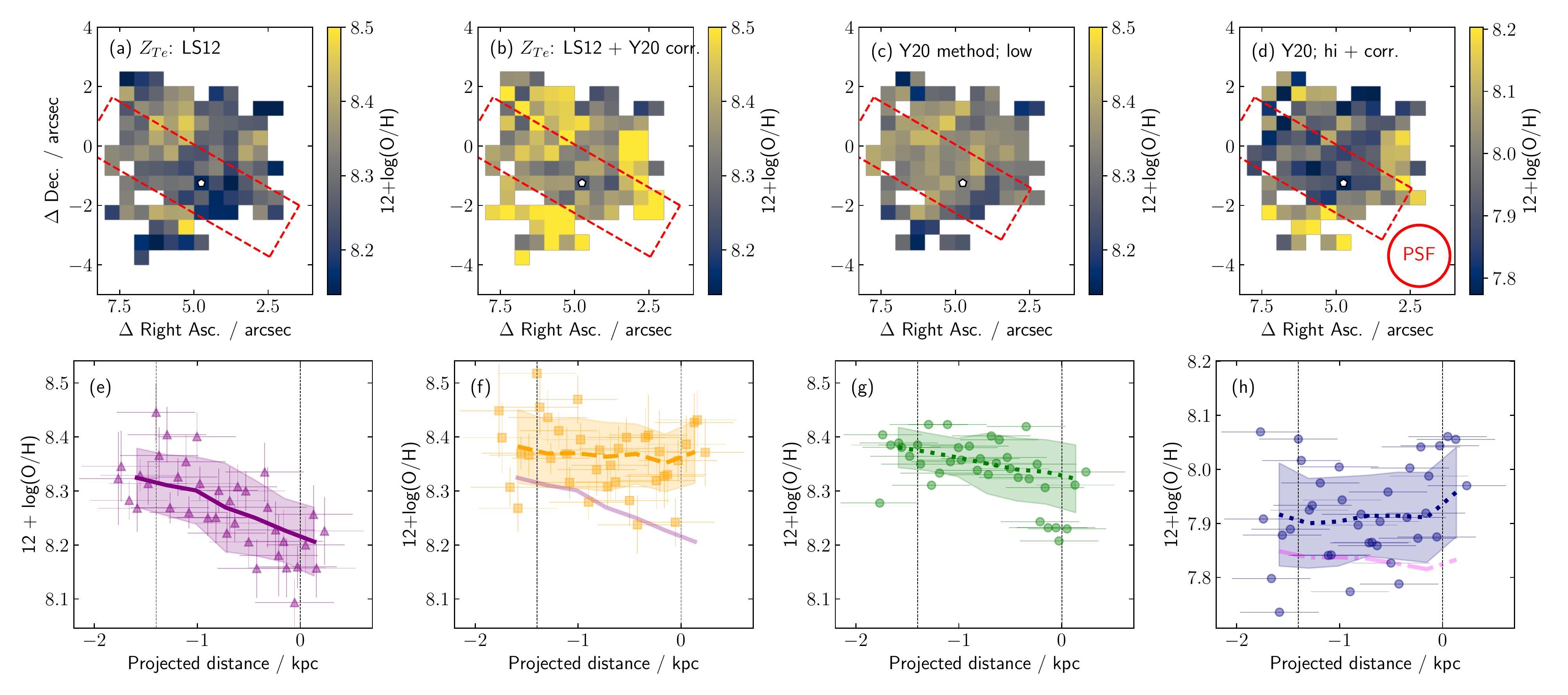}
    \caption{Comparison between the \ZLS method described in \S~\ref{sub:direct_logOH} and the two solutions arising from the \citetalias{Yates20} method. Far left and centre-left columns are simply reproduced from Figure~\ref{fig:direct_metal_map}.
    The centre-right column shows the \citetalias{Yates20} upper branch metallicities, while the far-right column shows corrected \citetalias{Yates20} lower branch metallicities (the uncorrected lower branch metallicity trend is shown in panel h as the magenta dot-dashed line).
    Note: We do not take the step of propagating measurement errors through the \citetalias{Yates20} method. However, the input measurement uncertainties are from the same source as those in panels (e) and (f) and thus would likely be comparable.}
    \label{fig:ap_y20_LS12_compare}
\end{figure*}

\subsection{Comparison between Yates method and LS12 method}
\label{ap:y20_3compare}

In line with our methods outlined in Appendix~\ref{ap:y20_2twoval} above, we consider two versions of the \citetalias{Yates20} method: one in which we adopt the ``upper branch'' solution to the \citetalias{Yates20} method, and another where we adopt the ``lower branch'' and apply the \citetalias{Yates20} empirical correction. We find that the resulting $T_e$(\OII) and $Z_\text{Te}$ maps are smooth with no unexpectedly large variations observed between pairs of adjacent spaxels.

In Figure~\ref{fig:ap_y20_LS12_compare} we compare the metallicities obtained from these \citetalias{Yates20} methods with those obtained in \S~\ref{sub:direct_logOH} from the \ZLS method. Metallicity maps and spatial trends in the far- and centre-left columns of Figure~\ref{fig:ap_y20_LS12_compare} are simply reproduced from Figure~\ref{fig:direct_metal_map} and show the \ZLS method with and without the \citetalias{Yates20} empirical correction.
The centre-right column shows the \citetalias{Yates20} upper branch metallicities, while the far-right column shows corrected \citetalias{Yates20} lower branch metallicities (additionally, the uncorrected lower branch metallicity trend is shown in panel h as the magenta dot-dashed line).

We first note that, even after the \citetalias{Yates20} correction, the lower branch metallicities are significantly lower than any of our other semi-direct methods.\footnote{And, indeed, strong-line methods; although some degree of offset is expected there \citep[e.g.][]{KewleyEllison08}} The large scatter in the values obtained makes it difficult to determine the spatial metallicity trend from this method, however, qualitatively it does seem to be broadly consistent with the trend seen in the \ZYates method applied in \S~\ref{sub:y20_correction}.

The normalisation of the upper branch values is in much better agreement with other semi-direct methods. The effect on the spatial trend is less clear: it appears somewhat flattened compared to \ZLS, however if a gradient were to be computed it would likely depend strongly on a cluster of lower metallicity points with projected distance $r\approx0$ kpc.

\citet{Yates20} showed that the semi-direct abundance deficit at low $O^{++}/O^+$ (which the \citetalias{Yates20} correction aims to address) is present across all $T_e$(\OII) -- $T_e$(\OIII) relations considered in that work, including LS12 (Figure 6 in \citealt{Yates20}). In particular, they show in detail its effect on the \citetalias{Yates20} lower branch metallicities.
Although it seems that low $O^{++}/O^+$ values seem to correlate with an increased preference for the upper branch metallicity solution, it is currently unclear whether variations in $O^{++}/O^+$ result in semi-direct metallicity biases in a similar way to that observed by \citet{Yates20} with respect to the lower branch solution and other $T_e$(\OII) -- $T_e$(\OIII) relations.
In the context of SAMI609396B where we have shown large scale variations in the \OIII/\OIIs ratio, the existence of such a bias could affect our interpretation of the spatial metallicity trend resulting from this upper branch solution.
Addressing this issue would require a more detailed analysis of a more extensive sample (e.g. the \citet{Yates20} sample) and is beyond the scope of this paper.

In summary, we find that the \citetalias{Yates20} semi-direct method as originally proposed favours a similar spatial gradient to that of \ZYates after the application of the \citetalias{Yates20} correction, albeit at a much lower normalisation. After identifying the two-valued nature of these relations, we found that adopting the upper branch values resulted in normalisation that agreed much better with other methods. We defer commenting on the spatial trend arising from these upper branch values to a later study, after the two-valued nature of the \citetalias{Yates20} relations has been examined in more detail.

Overall, these findings do not alter the main conclusion of this work: that assumptions around the temperature structure of \HII regions can have a significant impact on measured spatial metallicity trends in IFU observations of galaxies.


\bsp	
\label{lastpage}
\end{document}